%% file: main.tex
\documentclass[sigconf]{acmart}

\copyrightyear{2022} 
\acmYear{2022} 
\setcopyright{rightsretained} 
\acmConference[CCS '22]{Proceedings of the 2022 ACM SIGSAC Conference on Computer and Communications Security}{November 7--11, 2022}{Los Angeles, CA, USA}
\acmBooktitle{Proceedings of the 2022 ACM SIGSAC Conference on Computer and Communications Security (CCS '22), November 7--11, 2022, Los Angeles, CA, USA}
\acmDOI{10.1145/3548606.3560644}
\acmISBN{978-1-4503-9450-5/22/11}

\input{0-CCS2022-FinalVersion/commands}

\begin{document}
\title{Discovering IoT Physical Channel Vulnerabilities} 

\author{Muslum Ozgur Ozmen}
\affiliation{Purdue University}

\email{{mozmen@purdue.edu}}

\author{Xuansong Li}
\affiliation{School of Computer Science and Engineering, Nanjing University of Science and Technology}
\additionalaffiliation{State Key Laboratory for Novel Software Technology, Nanjing University}
\authornote{This work was completed while the authors were at Purdue University.}
\email{lixs@njust.edu.cn}
\authornote{Corresponding authors.}

\author{Andrew Chu}
\authornotemark[2]
\affiliation{University of Chicago}
\email{{andrewcchu@uchicago.edu}}

\author{Z. Berkay Celik} 
\authornotemark[3]
\affiliation{Purdue University}
\email{{zcelik@purdue.edu}}

\author{Bardh Hoxha}
\affiliation{Toyota Research Institute \\ North America}
\email{{bardh.hoxha@toyota.com}}

\author{Xiangyu Zhang}
\affiliation{Purdue University}
\email{xyzhang@cs.purdue.edu}

\begin{abstract}
Smart homes contain diverse sensors and actuators controlled by IoT apps that provide custom automation.
Prior works showed that an adversary could exploit physical interaction vulnerabilities among apps and put the users and environment at risk, \eg to break into a house, an adversary turns on the heater to trigger an app that opens windows when the temperature exceeds a threshold.
Currently, the safe behavior of physical interactions relies on either app code analysis or dynamic analysis of device states with manually derived policies by developers. 
However, existing works fail to achieve sufficient breadth and fidelity to translate the app code into their physical behavior or provide incomplete security policies, causing poor accuracy and false alarms.

In this paper, we introduce a new approach, \system, which efficiently combines app code analysis and dynamic analysis with new security policies to discover physical interaction vulnerabilities.
\system works by first translating sensor events and actuator commands of each app into a physical execution model (\apem) and unifying \apems to express composite physical execution of apps (\comp). 
\comp allows us to deploy \system in different smart homes by defining its execution parameters with minimal data collection.
\system supports new security policies with intended/unintended physical channel labels. It then efficiently checks them on the \comp via falsification, which addresses the undecidability of verification due to the continuous and discrete behavior of IoT devices.

We evaluate \system in an actual house with $14$ actuators, six sensors, and $39$ apps.
\system discovers $16$ unique policy violations, whereas prior works identify only $2$ out of $16$ with $18$ falsely flagged violations. 
\system only requires $30$ mins of data collection for each actuator to set the \comp parameters and is adaptive to newly added, removed, and relocated devices.
\end{abstract}



\maketitle
\renewcommand{\shortauthors}{Ozmen et al.}

\vspace{-3mm}

\input{0-CCS2022-FinalVersion/text/intro}
\input{0-CCS2022-FinalVersion/text/background}

\input{0-CCS2022-FinalVersion/text/motivation}

\input{0-CCS2022-FinalVersion/text/System_new}

\input{0-CCS2022-FinalVersion/text/eval-new}
\input{0-CCS2022-FinalVersion/text/discussion}
\input{0-CCS2022-FinalVersion/text/related}

\input{0-CCS2022-FinalVersion/text/conclusion}

\begin{acks}
We would like to thank Engin Masazade and Ali Cem Kizilalp for their feedback on the earlier version of this paper. 
This work has been partially supported by the National Science Foundation (NSF) under grants CNS-2144645, 1901242, and 1910300, DARPA VSPELLS under grant HR001120S0058, Rolls-Royce Cyber Technology Research Network Award, National Natural Science Foundation of China (No. 61702263), and the scholarship from China Scholarship Council (No. 201906845026). The views expressed are those of the authors only.
\end{acks}

\bibliographystyle{ACM-Reference-Format}
\bibliography{0-CCS2022-FinalVersion/bib/references}

\input{0-CCS2022-FinalVersion/text/appendix}

\end{document}

%% file: 0-CCS2022-FinalVersion/commands.tex
\usepackage{amsthm}
\usepackage{comment}
\usepackage{threeparttable}
\usepackage{multirow}
\theoremstyle{definition}

\makeatletter
\def\thm@space@setup{\thm@preskip=1pt
\thm@postskip=1pt}
\makeatother

\usepackage{graphicx}
\usepackage{algorithm}
\usepackage{amsmath}
\usepackage{algpseudocode}

\usepackage{tikz}
\usepackage{amsfonts}
\usepackage{ifthen}

\usepackage{enumitem}

\usepackage{mdframed}
\mdfsetup{skipabove=0.5pt,skipbelow=0.5pt}
\newmdenv{allfour}
\newmdenv[leftline=false,rightline=false, linecolor=darkgray, linewidth = 0.2mm, startinnercode={\baselineskip=0cm}]{topbot}
\newmdenv[topline=false,rightline=false]{leftbot}
\newmdenv[topline=false,bottomline=false]{leftright}

\usepackage{subcaption}
\usepackage{gensymb}
\usepackage{pifont}
\newcommand{\cmark}{\ding{51}}%
\newcommand{\xmark}{\ding{55}}%

\newboolean{commentsOn}
\setboolean{commentsOn}{true}

\ifthenelse{\boolean{commentsOn}}{
\newcommand{\berkay}[1]{{\color{purple}{\bf Berkay:} #1}}}
{\newcommand{\berkay}[1]{}}

\ifthenelse{\boolean{commentsOn}}{\newcommand{\xuansong}[1]{{\color{brown}{\bf LXS:} #1}}}
{\newcommand{\xuansong}[1]{}}

\ifthenelse{\boolean{commentsOn}}{\newcommand{\xz}[1]{{\color{red}[{\bf XZ:} #1]}}}
{\newcommand{\xz}[1]{}}

\ifthenelse{\boolean{commentsOn}}{\newcommand{\ozgur}[1]{{\color{blue}{\bf OO:} #1}}}
{\newcommand{\ozgur}[1]{}}

\ifthenelse{\boolean{commentsOn}}{\newcommand{\new}[1]{{\color{orange}#1}}}
{\newcommand{\new}[1]{}}

\ifthenelse{\boolean{commentsOn}}{\newcommand{\bardh}[1]{{\color{blue}{\bf BH:} #1}}}
{\newcommand{\bardh}[1]{}}

\ifthenelse{\boolean{commentsOn}}{\newcommand{\andrew}[1]{{\color{purple}{\bf AC:} #1}}}
{\newcommand{\andrew}[1]{}}

\ifthenelse{\boolean{commentsOn}}{
\newcommand{\raymond}[1]{{\color{green}{\bf Raymond:} #1}}}
{\newcommand{\raymond}[1]{}}

\usepackage{xspace}
\newcommand{\system}{{\textsc{\small{IoTSeer}}}\xspace}
\newcommand{\comp}{{\textsc{\small{CPeM}}}\xspace}
\newcommand{\apem}{{\textsc{\small{PeM}}}\xspace}
\newcommand{\apems}{{\textsc{\small{PeMs}}}\xspace}

\usepackage[scaled=.7]{beramono}

\DeclareRobustCommand*\circled[1]{\tikz[baseline=(char.base)]{ \node[shape=circle,draw,color=white,fill=black,inner sep=0.5pt] (char){#1};}}

\def\eg{{e.g.},~}

\usepackage{microtype}
\usepackage{amsmath,stackengine}
\stackMath
\usepackage{ifthen}

\newcommand{\squig}{{\scriptstyle\sim\mkern-3.9mu}}

\newcommand{\rsquigend}{{\scriptstyle\rule{.1ex}{0ex}\rhd}}
\newcounter{sqindex}
\newcommand\squigs[1]{%
  \setcounter{sqindex}{0}%
  \whiledo {\value{sqindex}< #1}{\addtocounter{sqindex}{1}\squig}%
}
\newcommand\rsquigarrow[2]{%
  \mathbin{\stackon[0.5pt]{\squigs{#2}\rsquigend}{\scriptscriptstyle\text{#1\,}}}%
}

\newcommand{\myrightarrow}[1]{\mathrel{\raisebox{-1pt}{$\mathtt{\xrightarrow{#1}}$}}}
\newcommand{\shortsectionBf}[1]{\vspace{2pt}
\noindent {\bf #1}}

\newcommand{\newTextBf}[1]{\vspace{1.5pt} \textbf{#1}}

\usepackage{setspace} 
\newcommand\staliro{{\textsc{\small{S-TaLiRo}}}\xspace}

\usepackage{url}
\usepackage{hyperref}
\mathchardef\mhyphen="2D
\usepackage{wasysym}

\makeatletter
\def\thickhline{%
  \noalign{\ifnum0=`}\fi\hrule \@height \thickarrayrulewidth \futurelet
   \reserved@a\@xthickhline}
\def\@xthickhline{\ifx\reserved@a\thickhline
               \vskip\doublerulesep
               \vskip-\thickarrayrulewidth
             \fi
      \ifnum0=`{\fi}}
\makeatother

\newlength{\thickarrayrulewidth}
\usepackage{listings}
\usepackage{xcolor}
\usepackage{color, colortbl}
\usepackage{tikz}
\definecolor{light-gray}{gray}{0.95}

\colorlet{punct}{red!60!black}
\definecolor{background}{HTML}{EEEEEE}
\definecolor{delim}{RGB}{20,105,176}
\colorlet{numb}{magenta!60!black}

\lstdefinelanguage{json}{
    basicstyle=\footnotesize\ttfamily,
    numbers=left,
    numberstyle=\scriptsize,
    stepnumber=1,
    numbersep=2pt,
    showstringspaces=false,
    breaklines=true,
    frame=lines,
    xleftmargin=1em,
    backgroundcolor=\color{background},
    literate=
     *{0}{{{\color{numb}0}}}{1}
      {1}{{{\color{numb}1}}}{1}
      {2}{{{\color{numb}2}}}{1}
      {3}{{{\color{numb}3}}}{1}
      {4}{{{\color{numb}4}}}{1}
      {5}{{{\color{numb}5}}}{1}
      {6}{{{\color{numb}6}}}{1}
      {7}{{{\color{numb}7}}}{1}
      {8}{{{\color{numb}8}}}{1}
      {9}{{{\color{numb}9}}}{1}
      {:}{{{\color{punct}{:}}}}{1}
      {,}{{{\color{punct}{,}}}}{1}
      {\{}{{{\color{delim}{\{}}}}{1}
      {\}}{{{\color{delim}{\}}}}}{1}
      {[}{{{\color{delim}{[}}}}{1}
      {]}{{{\color{delim}{]}}}}{1},
}

%% file: 0-CCS2022-FinalVersion/text/intro.tex
\section{Introduction}
\label{sec:intro}

With the growing number of IoT devices co-located in an environment, the interactions among IoT apps cause increasing safety and security issues~\cite{celik2019iotguard,manandhar2020towards,wang2019charting,zhang2019autotap,celik2019verifying,goksel2021safety}. 
There are two fundamental sources of app interactions, software and physical. 
Software interactions occur when IoT apps interact through a common device defined in their source code.
Consider an app that turns on the lights when smoke is detected and another app that locks the door when the lights are turned on. 
These apps interact through a common light device 
({\small{$\myrightarrow{smoke}$}} {\small{$\mathtt{light\mhyphen on}$}} {\small{$\myrightarrow{light\mhyphen on}$}} {\small{$\mathtt{door\mhyphen locked}$}}) and makes residents get trapped during a fire.

Physical interactions are another notable (and stealthier) threat; an app invokes an actuation command, and a sensor detects the physical channel influenced by this command, triggering other apps that actuate a set of devices.
Consider an app that turns on the heater and another app that opens the window when the temperature exceeds a threshold.
These apps interact through the temperature channel ({\small{$\mathtt{heater\mhyphen on}$}}$\mathbin{\stackon[-0.3pt]{\squigs{4}\rsquigend}{\scriptscriptstyle\text{temp.\,}}}$ {\small{$\mathtt{window\mhyphen open}$}}).
An adversary who exploits the heater controller app can stealthily trigger the {\small{$\mathtt{window\mhyphen open}$}} command and break into the house when the user is not home.

Discovering software and physical interactions has received increasing interest from the security community since they enable an adversary to indirectly gain control over sensitive devices and put the user and environment in danger.
Prior works mainly focus on identifying software interactions via app source code analysis~\cite{balliu2019securing2,balliu2020friendly,celik2019program,celik2018soteria,celik2019iotguard,chi2018cross,nguyen2018iotsan,surbatovich2017some}. 
These approaches find interacting apps by matching the device attributes in multiple apps, such as the {\small{$\mathtt{light\mhyphen on}$}} attribute in the first example. 
They cannot detect the physical interactions because the app source code does not state the physical channels, \eg the heater's influence on temperature.

There have been limited efforts to discover physical interactions. 
These works mainly ($1$) use pre-defined physical channel mappings between commands and sensor events~\cite{alhanahnah2020scalable} and ($2$) leverage NLP and device behavioral models to map the events and commands of apps~\cite{bu2018systematically,ding2018safety,wang2019charting}.
However, these approaches have limited expressiveness of physical channels, causing two issues.
They lead to over-approximation of physical channels, which are false alarms (\eg the system flags the temperature from {\small{$\mathtt{oven \mhyphen on}$}} and opens the window, yet the temperature from the oven is not enough to create a physical channel), and under-approximation of physical channels, which are the interactions that the system fails to identify (\eg the system ignores the motion from {\small{$\mathtt{robot \mhyphen vacuum \mhyphen on}$}}).

Recent work identifies physical interactions by collecting run-time device states and enforces security rules at run-time~\cite{ding2021iotsafe}.
However, this approach has three limitations, which limit its effectiveness. 
($1$) It defines rules based on devices' use cases to prevent physical interaction vulnerabilities. However, such rules do not prevent unintended interactions that subvert the intended use of apps and devices. For instance, a user installs an app that unlocks the patio door when motion is detected to automate their home entry process.
This system, however, does not prevent unlocking the patio door even if the motion is detected due to the vacuum robot's movements.
($2$) As a dynamic enforcement system, it cannot infer the specific command that influences a physical channel at run-time. For example, when the motion sensor detects motion, it cannot determine if this motion is from the vacuum robot or human presence.
($3$) When a device location changes, it makes wrong predictions about the app interactions, leading to unnecessarily enforcing rules and failing to prevent violations. For instance, if a portable heater is moved away from the temperature sensor, its influence on temperature measurements decreases. Yet, it would predict a higher influence and turn the heater off prematurely.

In this paper, we introduce \system, which builds the joint physical behavior of IoT apps through code and dynamic analysis, and validates a set of new security policies to discover physical interaction vulnerabilities.
\system first extracts an app's commands and sensor events from its source code. It translates them into physical execution models (\apems), which define each app's physical behavior.
It unifies the \apems of interacting apps in a composite physical execution model (\comp). 
To maximize \comp's fidelity in different smart homes, it collects device traces to define its execution parameters. 
\system supports new security policies that operate on intended/unintended physical channel labels and validates if IoT apps conform to these policies through falsification.
\system addresses the limitations of prior works with formal physical models of apps, new security policies, and validation through metric temporal logic. 

We applied \system in an actual house with $14$ actuators and six sensors, automated by $39$ apps from popular IoT platforms. 
We built the \apems of $24$ actuation commands and six sensor events used in the app source code and unified them in \comp. 
\system found $16$ unique physical interaction policy violations on different groups of interacting apps. 
We compared the violations discovered by \system with existing works that identify physical interaction vulnerabilities and found that they can only identify $2$ out of $16$ violations with $18$ false positives.
We repeated the experiments in the actual house and verified that all violations discovered by \system are true positives.
\system is adaptive to newly added, removed, and updated devices and imposes minimal model construction and validation time overhead.
It requires, on average, $30$ mins of data collection for each actuator to set the parameters, and it takes, on average, $21$ secs to validate a physical channel policy on four interacting apps. 

In this paper, we make the following contributions.
\begin{itemize}
\item \textbf{Translating App Source Code into its Physical Behavior}:  We translate the actuation commands and sensor events in the app source code into physical execution models to define their physical behavior.

\item \textbf{Composition of Interacting Apps}: We introduce a novel composite physical execution model architecture that defines the joint physical behavior of interacting apps. 

\item \textbf{Physical Channel Policy Validation}: We develop new security policies with intended/unintended physical channel labels.
We formally validate the policies on \comp through optimization-guided falsification.

\item \textbf{Evaluation in an Actual House}: We use \system in a real house containing $14$ actuators and six sensors and expose $16$ physical channel policy violations.

\item 
\system code is available at 
\vspace{-2pt}
\begin{center}
    \url{https://github.com/purseclab/IoTSeer}
\end{center}
\vspace{-2pt}
for public use and validation.
\end{itemize}

%% file: 0-CCS2022-FinalVersion/text/background.tex
\section{Motivation and Threat Model}
\label{sec:background}
A smart home is composed of a set of apps that monitor and control sensors and actuators.
Apps subscribe to events (\eg {\small{$\mathtt{motion \mhyphen}$}} {\small{$\mathtt{detected}$}}) that invoke their event handler methods, activating actuation commands (\eg {\small{$\mathtt{door \mhyphen unlock}$}}).
Users install official apps from IoT markets such as HomeKit~\cite{HomeKitWebsite} and OpenHAB~\cite{OpenHabWebsite}, and third-party apps through proprietary web interfaces.
Another trend for custom automation is trigger-action platforms such as IFTTT~\cite{IFTTTWebsite} and Zapier~\cite{ZapierWebsite}. 
These platforms allow users to use conditional statements in the form of if/then rules to integrate digital services with IoT devices.
In this paper, we use the term app(s) to refer to both IoT apps and trigger-action rules.

%% file: 0-CCS2022-FinalVersion/text/motivation.tex
When an actuation command is invoked, it influences a set of physical channels measured by sensors. The command then interacts with apps subscribed to those sensor events, invoking other commands. 
An adversary can exploit such physical interactions to indirectly control devices and cause unsafe states.

\begin{figure}[t!]
    \centering
    \includegraphics[width=\linewidth]{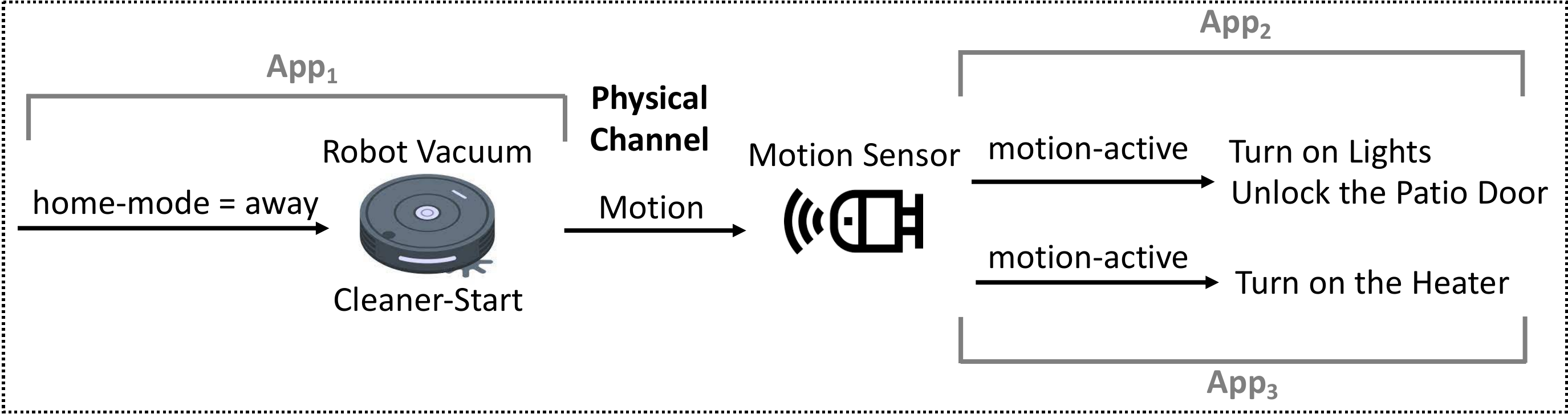}
    \caption{Illustration of physical interactions: the vacuum robot is activated when the user is not home, which unlocks the patio door and turns on the lights and heater.}
    \label{fig:motivation}
\end{figure}

To illustrate, Figure~\ref{fig:motivation} shows a common smart home with four devices.
The user installs {\small{$\mathtt{App_1}$}} that starts the robot vacuum cleaner when the home mode is set to away (or at specific times).
An adversary can provide users with {\small{$\mathtt{App_2}$}} and {\small{$\mathtt{App_3}$}} that operate correctly in isolation yet exploit the physical interactions to cause unsafe states.
When motion is detected (the user enters the home), {\small{$\mathtt{App_2}$}} turns on the lights and unlocks the patio door, and {\small{$\mathtt{App_3}$}} sets the heater to a specific temperature value.

In this deployment, the user leaves home and sets the home mode to away, triggering {\small{$\mathtt{App_1}$}} that starts the robot vacuum cleaner.
The movements of the robot vacuum create a physical interaction with {\small{$\mathtt{App_2}$}} and {\small{$\mathtt{App_3}$}} since the motion sensor detects the robot vacuum.
This results in unlocking the patio door and turning on the lights and the heater while the user is not at home. 
The unlocked patio door may allow a burglar to break in, the turned-on lights may indicate whether the users are at home or not, and the heater's influence on temperature may trigger other apps (\eg opening the windows), causing a chain of interactions between multiple apps.

The preceding example shows that the final environment states do not just depend on individual devices but are a result of the physical interactions of multiple devices. 
Each app is individually safe, yet their unified physical interactions leave users at risk. 

\subsection{Threat Model}
\label{sec:threat}
Our threat model is similar to related IoT security works, which focus on app interaction vulnerabilities~\cite{ding2021iotsafe,ding2018safety,wang2019charting, celik2019iotguard}. 
We consider an adversary whose goal is to execute undesired device actions (\eg unlocking the door when the user is sleeping) and cause unsafe system states. 
The adversary achieves this goal by creating or exploiting physical app interactions.
The adversary can conduct two types of attacks, a mass attack or a targeted attack.
In a mass attack, an adversary provides users with apps operating correctly in isolation yet exploits the physical interactions among apps to cause unsafe states on a large scale. 
The adversary does not target a specific smart home but harms many users and hurts the trustworthiness of an IoT platform.
The adversary can conduct this attack by ($1$) distributing apps on IoT platforms and third-party IoT forums and ($2$) tricking users into installing apps via phishing and other social engineering methods.

In a targeted attack, the adversary determines a specific smart home to exploit its physical app interactions.
First, the adversary discovers an exploitable physical interaction in the target smart home.
For this, the adversary remotely learns the devices and installed apps by eavesdropping on the commands and sensor events over network packets and mining their correlations~\cite{fu2021hawatcher,acar2018peek}.
The adversary can then wait until the physical interactions naturally occur and create unsafe states (\eg the door is unlocked when the user is not at home) to conduct a physical attack. 
The adversary can also leverage vulnerable apps to remotely control a set of commands and cause physical interactions~\cite{ding2018safety,ding2021iotsafe,wang2019charting}.
Through this, the adversary can stealthily invoke actuation commands through physical channels even if they cannot directly control them.

The physical interactions might also happen due to the errors in users' creation, installation, and configuration of apps. 
In such cases, the physical interactions subvert the intended use of IoT devices, leading to unsafe states. 
This is because IoT users are usually uninformed about the implications of app interactions, as demonstrated by prior works~\cite{ur2016trigger,zhao2020visualizing}.

\begin{figure*}[ht!]
    \centering
    \includegraphics[width=\textwidth]{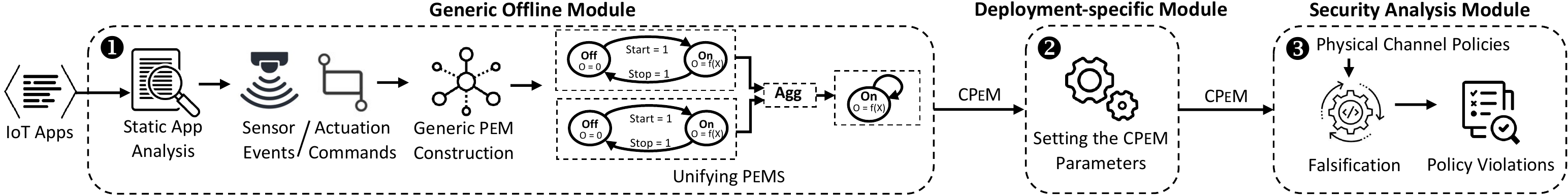}
    \caption{Overview of \system's architecture}
    \label{fig:overview}
\end{figure*}

\section{Design Challenges}
\label{sec:challenges}

\shortsectionBf{C1: Correct Physical Interactions.}  
To identify physical interaction vulnerabilities, prior works have used NLP techniques, \eg the heater is semantically related to temperature~\cite{ding2018safety}, manually crafted interaction mappings, \eg {\small{$\mathtt{heater \mhyphen on}$}} is mapped to the temperature channel~\cite{alhanahnah2020scalable}, and constructed naive device models, \eg {\small{$\mathtt{heater \mhyphen on}$}} increases temperature by $\mathtt{1\degree C}$ in $\mathtt{8}$ hours~\cite{wang2019charting}.

These approaches, unfortunately, discover erroneous interactions or fail to discover them due to over-approximating and under-approximating physical channel properties. 
Such errors may cause serious consequences. For instance, when the user is not at home, they fail to block {\small{$\mathtt{door \mhyphen unlock}$}}  or mistakenly approve  {\small{$\mathtt{window \mhyphen open}$}}.

\shortsectionBf{C2: Unintended Physical Interactions.}  
Prior works define security rules based on the use cases of devices to prevent physical interaction vulnerabilities.
Such rules do not consider unintended interactions, which occur beyond the intended use of devices and apps, and unexpectedly trigger actions in a smart home. %
For instance, the user uses {\small{$\mathtt{App_2}$}} and {\small{$\mathtt{App_3}$}} in Figure~\ref{fig:motivation} to turn on the light and heater and unlock the patio door when they enter the home.
However, when the vacuum robot creates motion, it unintentionally triggers both of these apps and invokes their actions.

Additionally, an actuation command may unintentionally trigger a security rule and subvert its intended use. 
For example, prior works define a rule that states, ``The alarm must sound and an SMS/Push message should be sent to the owner when motion is detected, and home mode is away'' to protect the smart home from intruders~\cite{ding2021iotsafe}.
This rule can be triggered when {\small{$\mathtt{App_1}$}} turns on the vacuum robot, which would create panic and unnecessarily bring resources (\eg police dispatch) to the home.

\shortsectionBf{C3: Run-time Dilemmas.}
Dynamic systems that examine the device states at run-time~\cite{celik2019iotguard,ding2021iotsafe} cannot infer the influence of an exact command on a physical channel. 
For instance, in Figure~\ref{fig:motivation}, it is unclear to these systems whether the motion is from the vacuum robot or human presence.
This challenge becomes more critical when multiple devices influence the same physical channel. 
For example, if a sound sensor detects the sound from both AC and dryer, the dynamic systems cannot determine if a single device or their aggregated influence changes the sound. 
This makes them flag incorrect physical interactions.

When an interaction vulnerability is identified, dynamic systems either block device actions or notify users. However, these responses could be dangerous. 
For example, the door-unlock action might be blocked if there is a fire in the house when the user is not home.

\shortsectionBf{C4: Device Placement Sensitivity.}  
Prior works do not model the impact of the distance between an actuator and a sensor on physical interactions.
Intuitively, if the distance between an actuator and sensor increases, the physical influence of a command on sensor readings decreases monotonically. 
Thus, when a device's placement is changed, the identified interactions may no longer occur, and there may be new interactions that were not previously identified. 

This observation causes wrong predictions with false positives (incorrect policy violations and unnecessarily enforcing policies) and false negatives (missing violations and failing to prevent them).
This is critical in smart homes as frequent device placement changes may occur with lightweight and portable IoT devices.

%% file: 0-CCS2022-FinalVersion/text/System_new.tex
\section{IoTSeer Design}
\label{sec:system}

To discover physical interaction vulnerabilities, we introduce \system, which combines app code analysis and dynamic analysis with new security policies, and efficiently addresses the \textbf{C1}-\textbf{C4} challenges. Figure~\ref{fig:overview} provides an overview of \system's modules.

In the \emph{generic offline} module (\circled{\small{1}}), \system first extracts actuation commands and sensor events of apps from their source code via static analysis.
From this, it builds \emph{physical execution models} (\apems) for each physical channel a command influences and a sensor measures.
Each \apem defines a generic physical behavior of commands and events in hybrid automata with well-studied generic differential and algebraic equations.

\system then unifies the \apems in a \emph{composite physical execution model} (\comp) to represent the joint physical behavior of interacting apps. 
Our composition algorithm considers a set of physical channel properties (\eg the aggregation of physical influences) (\textbf{C1}) and distinguishes the influence of each command (\textbf{C3}).

The offline module delivers \apems and \comp that define a generic physical behavior for devices and their composition on the physical channels. 
However, each smart home may contain devices with different properties (\eg heater power) and environmental factors (\eg furniture and room layout).

To address this, in the \emph{deployment-specific} module (\circled{\small{2}}), \system extends RSSI-based localization~\cite{adewumi2013rssi,zanca2008experimental} to obtain the physical distance between actuators and sensors.
It next collects device traces and leverages system identification techniques to define the execution parameters of \comp.
The resulting \comp defines the physical behavior of interacting apps for a specific smart home with minimal data collection and addresses the device location changes (\textbf{C4}).

In the security analysis module (\circled{\small{3}}), we first develop security policies to detect unintended physical interactions that cause unsafe and undesired system states (\textbf{C2}).
\system then extends optimization-guided falsification to validate if the joint physical behavior of interacting apps conforms to the identified policies.
If \system discovers a policy violation, it outputs the violation's root cause with the interacting apps and physical channels.

\begin{figure}[t!]
    \centering
    \includegraphics[width=.8\linewidth]{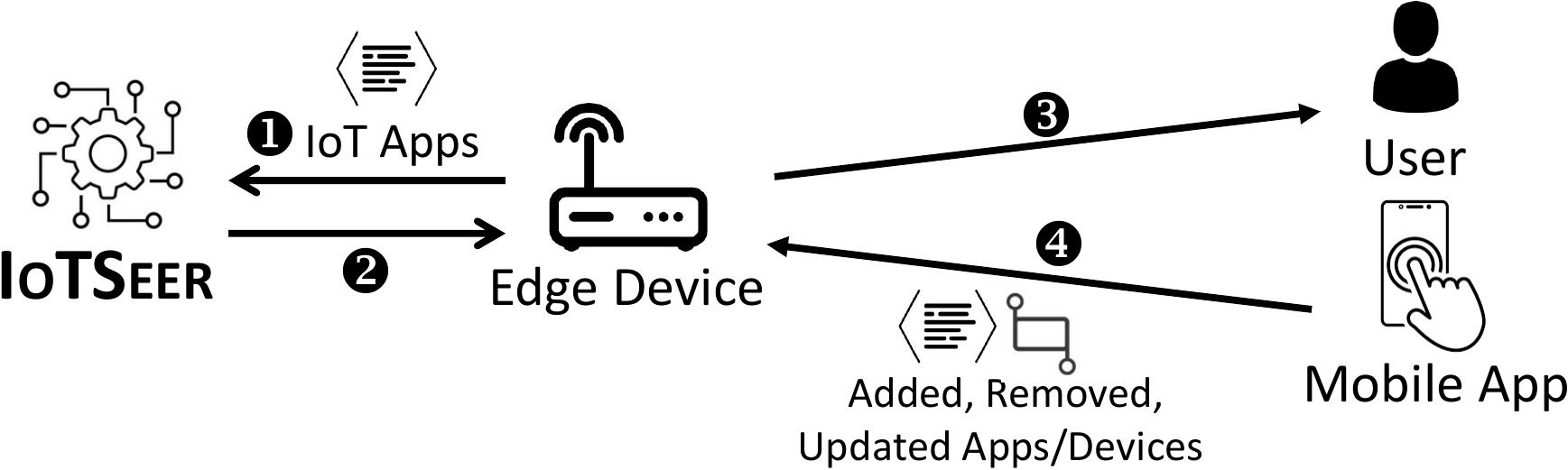}
    \caption{Usage scenario of \system} 
    \label{fig:usage}
\end{figure}

\shortsectionBf{Deployment.}
Our \system prototype runs in conjunction with the edge device in a smart home (See Figure~\ref{fig:usage}). However, it could be implemented as a software service in the cloud or in a local server.

\system first obtains the IoT apps and runs its generic offline module (\circled{1}).
It then collects actuator and sensor traces through the edge device for the deployment-specific module (\circled{2}).
It next runs its security analysis module and presents users with the policy violations and their root causes (\circled{3}).

\system supports dynamic changes in the smart home, including added, removed, and updated IoT apps and devices (\circled{4}). 
When a dynamic change occurs, \system reruns its related modules and presents users with the changed policy violations and their root causes.
First, if the user installs a new app or device, \system runs the generic offline and deployment-specific modules to include the new apps and devices in the \comp. 
Second, if the user removes an app or device, \system removes their \apems and transitions from the \comp.
Lastly, if an app's configuration or a device's placement changes, \system changes the \comp's parameters with the deployment-specific module. 
After updating the \comp, \system runs the security analysis module to identify policy violations. 

\begin{figure*}[t!]
    \centering
    \includegraphics[width=\textwidth]{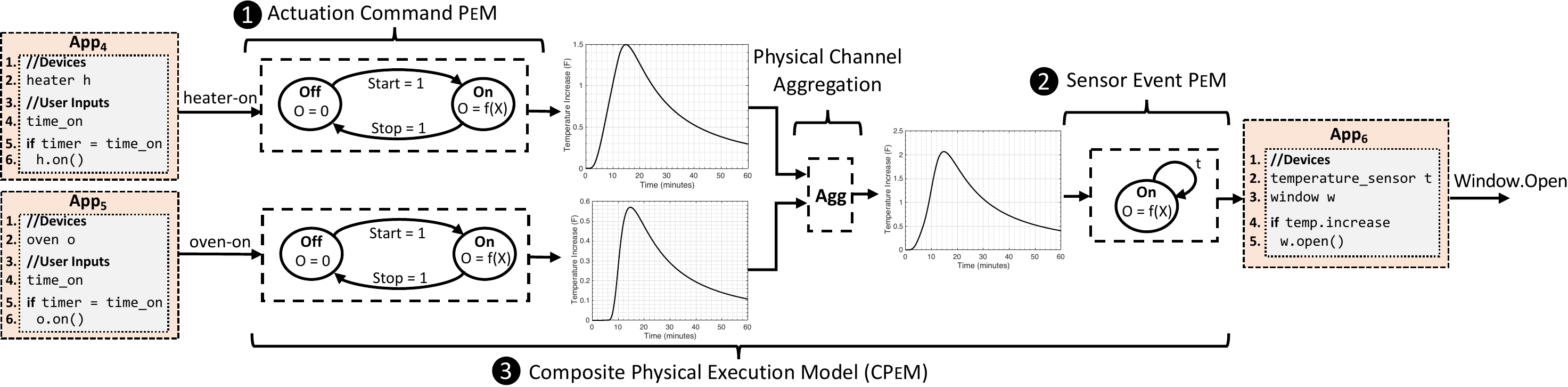}
    \caption{Illustration of \apems for actuators and sensors, and their \comp for the unified behavior of three apps ({\small{$\mathtt{App_4}$}}, {\small{$\mathtt{App_5}$}}, {\small{$\mathtt{App_6}$}}).}
    \label{fig:automata2}
\end{figure*}

\subsection{Generic Offline Module}

To map an IoT app source code to its physical behavior, \system requires an app's events, actuation commands, and trigger conditions associated with each command. 
However, IoT platforms are diverse, and each offers a different programming language for automation. For instance, IoT platforms such as OpenHAB enable users to write apps with a Domain Specific Language based on Xbase~\cite{OpenHabWebsite}, and trigger-action platforms such as IFTTT implement if-then abstractions~\cite{IFTTTWebsite}. 
To address this, we leverage existing static analysis and parsing tools for IoT apps~\cite{celik2019program,celik2018soteria,wang2019charting}.

In this way, \system supports apps from various IoT platforms.
These tools model an app’s life-cycle, including its entry points and event handlers from its interprocedural control flow graph (ICFG), and extract ($1$) devices and events, ($2$) actuations to be invoked for each event, and ($3$) conditions to invoke the actuations.
For instance, given an app \emph{``When the temperature is higher than $80 \degree F$, if the AC is off, then open the window''}, \system obtains the {\small{$\mathtt{temp > 80}$}} event, the {\small{$\mathtt{window \mhyphen open}$}} command, and the {\small{$\mathtt{AC \mhyphen off}$}} trigger condition. 

\subsubsection{Constructing \apems}
\label{subsec:constructing}

We translate each command and sensor event of an app to a \apem expressed with a hybrid I/O automaton.
This process begins by constructing a separate \apem for each physical channel a command influences, and a sensor event observes with \emph{physics-based modeling}.
The physics-based modeling integrates a generic differential or algebraic equation from control theory into a \apem to model each app's physical behavior~\cite{jackson1996sensor,voudoukis2017inverse,zhivov2001principles}.
This approach is widely used in robotic vehicles~\cite{ArdupilotSITL,PX4SITL} (\eg to predict RV's sensor values) and in autonomous vehicles~\cite{Carla_Physics} (\eg to model the movements of cars and pedestrians).

Determining the physical channels that a command influences requires collecting actuator and sensor traces from the smart home since static app analysis does not reveal the commands' physical channels. \system initially considers each command may influence all physical channels and removes the over-approximated channels in the deployment-specific module (See Sec.~\ref{subsec:tuning}).

\shortsectionBf{\apems for Actuation Commands.}
A command \apem defines the discrete and continuous dynamics of a command. 
The discrete behaviors are an actuator's states (\eg on/off) for invoking the command from the app.
The continuous behavior is an algebraic or differential equation that defines its physical behavior.

Formally, each \apem is a hybrid I/O automaton~\cite{lynch2003hybrid} in the form of {\small{$\mathtt{H_a} = (\mathtt{Q}$, $\mathtt{X}$, $\mathtt{f}$, $\mathtt{\rightarrow}$, $\mathtt{U}$, $\mathtt{O}$)}}. 
Here  {\small{$\mathtt{Q}$}} is a set of discrete states, {\small{$\mathtt{X}$}} is a continuous variable, {\small{$\mathtt{f}$}} is a flow function that defines the continuous variable's evolution, ({\small{$\mathtt{\rightarrow}$}}) defines the discrete transitions, and {\small{$\mathtt{U}/\mathtt{O}$}} defines the input/output variables, as shown in Figure~\ref{fig:automata2}-\circled{1}.
We define the discrete states as {\small{$\mathtt{Q = \{on, off\}}$}}, and discrete transitions enable switching between them. 
The continuous variable defines a command's influence on physical channels (\eg temperature in {\small{$\mathtt{\degree F}$}}, sound in {\small{$\mathtt{dB}$}}). The flow function acts on the continuous variable, and the \apem outputs the command's influence. 

We define a separate generic flow function for each physical channel.
They are \emph{differential} equations for continuous physical channels such as temperature and \emph{algebraic} equations for instant channels such as sound.
A flow function takes two parameters as input, device property, and distance from the actuator and outputs the actuator's influence on a physical channel at that distance.
These parameters allow us to use the same flow function for different actuators that influence the same channel (\eg {\small{$\mathtt{heater \mhyphen on}$}} and {\small{$\mathtt{AC\mhyphen on}$}}) and the actuators with multiple working patterns (\eg {\small{$\mathtt{AC}$}}'s modes) by setting different parameters.

The property parameter describes the characteristics of a device, such as its operating power. 
In Sec.~\ref{subsec:deployment}, we show how to set the parameters based on a specific smart home with \system's deployment-specific module for precision.
The distance parameter quantifies the command's influence at different locations (\eg {\small{$\mathtt{fan\mhyphen on}$}}'s sound intensity at $1$ and $2$ meters away from the fan). 
We set this parameter as the distance from the actuator to the sensor that measures its influence (Sec.~\ref{subsec:deployment}). 
This makes the \apem practical against device placement changes and enables effortless porting of \system to other deployments with different placements.

\newTextBf{Example Actuator \apem.} 
We illustrate a \apem for actuators that influences the temperature channel.
The flow function for commands that influence temperature uses the partial differential heat diffusion equation~\cite{hancock20061}, $(\partial \mathtt{T}) / (\partial \mathtt{t}) = \alpha(\partial^2\mathtt{T}) / (\partial \mathtt{x}^2)$ with boundary conditions.
Here, $\mathtt{T}$ is the environment's temperature ($\degree \mathtt{K}$), $\mathtt{x}$ is the distance parameter ($\mathtt{m}$), $\alpha$ is the thermal diffusivity constant ($\mathtt{m}^2/\mathtt{s}$), and the boundary conditions define the actuator's temperature.  

Given the device property parameter and the distance to the temperature sensor, the \apem outputs the command's influence on the temperature sensor's measurements over time.

\shortsectionBf{\apems for Sensor Events.} 
We define an event's \apem as a hybrid I/O automaton ({\small{$\mathtt{H_s}$}}) with a single state, {\small{$\mathtt{Q = \{on\}}$}}, and a timed ({\small{$\mathtt{t}$}}) self transition {\small{$\mathtt{on \myrightarrow{t} on}$}}, where {\small{$\mathtt{t}$}} is the frequency that a sensor samples its measurements.
Figure~\ref{fig:automata2}-\circled{2} depicts a sensor event's \apem that only measures physical channels.

The sensor event \apem takes a sensitivity-level parameter, which defines the minimum amount of change in the physical channel (threshold) for a sensor to change its reading.
We set the sensitivity level based on the sensors installed in the smart home.
A threshold function outputs a sensor reading indicating if the physical channel level is equal to or greater than the sensitivity level.
If the sensor measures boolean-typed values (\eg motion), the \apem outputs a bit indicating ``{\small{$\mathtt{detected}$}}'' or ``{\small{$\mathtt{undetected}$}}'' events.
If the sensor makes numerical readings (\eg temperature), it outputs numerical values.

\newTextBf{Example Sensor \apem.} 
We illustrate a \apem for a sound sensor that outputs boolean-typed measurements.
The threshold function of sound sensor events is defined as {\small{$\mathtt{f(sp) = 1~if~sp > th, 0}$}} {\small{$\mathtt{otherwise.}$}} where {\small{$\mathtt{sp}$}} is the ambient sound pressure and {\small{$\mathtt{th}$}} is the sensor's threshold (sensitivity level).
Here, the \apem outputting $\mathtt{1}$ means ``{\small{$\mathtt{sound \mhyphen detected}$}}'' and $\mathtt{0}$ means ``{\small{$\mathtt{sound \mhyphen undetected}$}}.''

\shortsectionBf{Built-in \apems.}
Using the above approach, we have integrated into \system a total of $24$ actuator command \apems (\eg {\small{$\mathtt{heater \mhyphen on}$}}, {\small{$\mathtt{door \mhyphen unlock}$}}) that influence a total of six physical channels, namely temperature, humidity, illuminance, sound, motion, and smoke, and six sensor event \apems that measure these channels. 
The \apems can be easily extended to define the physical behavior of various devices since their flow functions are generic for a family of devices that influence the same physical channel.

The \apems allow us to obtain the physical behavior of popular apps used in diverse IoT platforms. 
We detail their hybrid I/O automata in Appendix~\ref{sec:indModelFormulae} and evaluate them in our evaluation in Sec.~\ref{sec:eval}.

\subsubsection{Unifying the Physical Behavior of Apps}

After we build the \apems for sensors and actuators to define the behavior of each app, we build a separate \comp to represent their joint behavior.

Algorithm~\ref{alg:composition} presents our approach to \comp construction.
The algorithm starts with identifying the interacting apps by matching the physical channels of sensor events and commands. 
First, if a sensor measures a physical channel that a command influences, we add a transition from the command \apem ({\small{$\mathtt{H_a}$}}) output to the sensor event \apem ({\small{$\mathtt{H_s}$}}) input (Lines 2-4).
Second, software and physical channels can trigger the event handler of apps and invoke commands if the apps' conditions are satisfied. 
For physical channels, we add a transition from a sensor event \apem ({\small{$\mathtt{H_s}$}}) to a command \apem ({\small{$\mathtt{H_a}$}}) (Lines 5-12).
For software channels, we add a transition from a command \apem ({\small{$\mathtt{H_{a_1}}$}}) to another command \apem ({\small{$\mathtt{H_{a_2}}$}}) if an app invokes {\small{$\mathtt{a_2}$}} when {\small{$\mathtt{a_1}$}} occurs (Lines 13-17).
The transitions are expressed with a {\small{$\mathtt{UNIFY}$}} operator, which defines the interactions as a transition, {\small{$\mathtt{H_a {\rsquigarrow{}{2}} H_s}$}}, {\small{$\mathtt{H_s {\myrightarrow{}} H_a}$}}, and {\small{$\mathtt{H_a {\myrightarrow{}{}} H_a}$}}. 
Here, {\small{$\mathtt{{\rsquigarrow{}{2}}}$}} is a physical influence on a channel, and {\small{$\mathtt{{\myrightarrow{}}}$}} is a software channel. 

\input{0-CCS2022-FinalVersion/algorithms/composition}

Figure~\ref{fig:automata2}-\circled{3} shows the \comp of three apps ({\small{$\mathtt{App_4}$}}, {\small{$\mathtt{App_5}$}}, {\small{$\mathtt{App_6}$}}) that automate a heater, oven, window, and temperature sensor.
When {\small{$\mathtt{App_4}$}} invokes {\small{$\mathtt{heater \mhyphen on}$}} and {\small{$\mathtt{App_5}$}} invokes {\small{$\mathtt{oven \mhyphen on}$}}, \system identifies {\small{$\mathtt{heater \mhyphen on}$}}'s and {\small{$\mathtt{oven \mhyphen on}$}}'s temperature \apems interact with temperature sensor of {\small{$\mathtt{App_6}$}}.
\system adds the below transitions to the \comp: 

\vspace{1mm}
\begin{topbot}
\vspace{-1mm}
\hspace{-3mm} {\footnotesize{$\mathtt{H_a\{ heater \mhyphen on\}}$}}
$\mathbin{\stackon[-0.3pt]{\squigs{2}\rsquigend}{\scriptscriptstyle\text{\,}}}$
{\footnotesize{$\mathtt{H_s\{temp \mhyphen increase\}}$}}

\hspace{-5.5mm}{\footnotesize{$\mathtt{H_a\{ oven \mhyphen on\}}$}}
$\mathbin{\stackon[-0.3pt]{\squigs{2}\rsquigend}{\scriptscriptstyle\text{\,}}}$
{\footnotesize{$\mathtt{H_s\{temp \mhyphen increase\}}$}}
\vspace{-.5mm}
\end{topbot}
\vspace{1mm}

Another transition from the temperature sensor event \apem to {\small{$\mathtt{window \mhyphen open}$}} \apem is then added because when the sensor measures an increased temperature, {\small{$\mathtt{App_6}$}} opens the window. 

\vspace{1mm}
\begin{topbot}
\vspace{-1mm}
\hspace{-3mm} {\footnotesize{$\mathtt{H_s\{temp \mhyphen increase\}}$}}
{\footnotesize{$\myrightarrow{}$}} 
{\footnotesize{$\mathtt{H_a\{window \mhyphen open}\}$}}
\vspace{-.5mm}
\end{topbot}
\vspace{1mm}

\shortsectionBf{Addressing Aggregation and Dependency.}
A sensor measures the accumulated influence of multiple commands. 
For this, we define an aggregation operator ({\small{$\mathtt{AGG}$}}), which combines {\small{$\mathtt{UNIFY(H_a,H_s)}$}} operators so that a sensor event \apem takes the aggregated output of command \apems as input (Lines 18-19). 
Turning to Figure~\ref{fig:automata2}, \system adds {\small{$\mathtt{Agg(H_{a}^1\{heater \mhyphen on\}}$}}, {\small{$\mathtt{H_{a}^2\{oven \mhyphen on \}}$}}{\small{$\mathtt{)}$}}  {\small{${\mathtt{\mathbin{\stackon[-0.3pt]{\squigs{2}\rsquigend}{\scriptscriptstyle\text{\,}}}}}$}} {\small{$\mathtt{temp \mhyphen increase}$}} transition to the \comp.
The {\small{$\mathtt{AGG}$}} operator's output is defined based on the physical channel's unit.
It is the sum of the command \apem outputs, {\small{$\mathtt{\sum_{i=1}^{n} UNIFY(H_a^{i}, H_s)}$}}, for linear scale channels (\eg temperature)~\cite{zhivov2001principles}.
The channels in the log scale (\eg sound) are aggregated after being converted to a linear scale, {\small{$\mathtt{10 \times log_{10}(\sum_{i=1}^{n} 10^{H_a^{i}/10})}$}}~\cite{mitschke2009decibel}.

Another property of physical channels is that a physical channel ({\small{$\mathtt{p_j}$}}) may depend on another channel ({\small{$\mathtt{p_i}$}}) if a change in {\small{$\mathtt{p_i}$}} affects {\small{$\mathtt{p_j}$}}. 
Due to dependencies, a sensor event \apem's output may influence another sensor event \apem's readings. 
The generic \apems allow us to easily identify dependencies by iteratively taking each sensor event \apem and checking if it is used in the threshold function of another sensor event (Lines 20-23). 
For instance, when ambient temperature increases, the air-water capacity increases and affects the humidity sensor's readings~\cite{lawrence2005relationship}. 
To address this, we add a {\small{$\mathtt{DEP(H_{s}^i \rightarrow H_{s}^j)}$}} transition from the temperature sensor event \apem output ({\small{$\mathtt{H_{s}^i}$}}) to the humidity sensor event \apem input ({\small{$\mathtt{H_{s}^j}$}}).

\subsection{Deployment-specific Module}
\label{subsec:deployment}

The actuation command \apems require distance and device property parameters, and the sensor event \apems require a sensitivity level parameter.
We set these parameters based on the devices installed in the smart home to ensure the \comp precisely models the physical behavior of the apps.

\subsubsection{Setting the Distance Parameter}
\label{subsec:tuning_distance}
To determine the distance parameter in \apems, we initially considered leveraging a recent IoT device localization tool, Lumos~\cite{sharma2022lumos}.
Lumos localizes IoT devices with high accuracy by requiring the user to walk around the smart home with a mobile phone.
However, this approach could be inconvenient for smart home users since it requires manual effort.
To address this, we integrate received signal strength intensity (RSSI)-based distance estimation techniques~\cite{adewumi2013rssi,zanca2008experimental} into \system.
Such techniques leverage the inverse proportion between distance and RSSI to estimate the distance between two devices.
Although this approach may incur an error in the distance parameter, our evaluation shows that the impact of such errors on \system's policy violation identification is minimal (See Sec.~\ref{sec:eval}).

\subsubsection{Setting the Device Property Parameter}
\label{subsec:tuning}
We consider two options for setting the device property parameters based on the installed devices.
The first is using the installed device's datasheets. 
Yet, in our prototype implementation, we realized that the datasheets might be incomplete, or a discrepancy could occur, for example, due to device aging~\cite{refregier2004noise}. 
To address this, we extend System Identification (SI), a learning-based method commonly used by control engineers to estimate parameters or models of physical processes using experimental data traces~\cite{keesman2011system,Ptolemaeus}. 

\input{0-CCS2022-FinalVersion/tables/policies}

SI allows us to estimate device property parameters that ensure the \comp achieves high fidelity with actual devices.
This process requires \emph{fewer traces} than the traditional application of SI as we only estimate parameters instead of complete equations (See Sec.~\ref{sec:eval}).
We particularly use $(\tau,\epsilon)$-closeness~\cite{abbas2014formal} as the fidelity metric.
$(\tau,\epsilon)$-closeness determines the difference between two traces in their timing ($\tau$) and values ($\mathtt{\epsilon}$), where $\mathtt{\epsilon}$ is referred to as deviation score.

To apply this approach in a smart home, \system individually activates each actuator and collects sensor measurements. 
It next runs their \apems with a device property parameter and obtains sensor traces.
It computes the $(\tau,\epsilon)$-closeness between the actual device and \apem traces.
It conducts a binary search on the device property parameter to obtain the optimal value that minimizes the deviation score. 
Using real device traces to determine the device property parameters ensures that the impact of environmental conditions (\eg furniture) on sensor readings is integrated into the \comp.

From the collected actual device traces, \system also determines the set of physical channels that a command influences in the smart home.
\system checks if a command does not change the sensor measurements or if its influence is statistically indistinguishable from environmental noise.
In such a case, \system removes the \apems of those commands from the \comp. 

\subsection{Security Analysis Module}
We first identify intent-based policies to detect unintended app interactions and device-centric policies to detect the vulnerabilities intended interactions present (Sec.~\ref{subsec:policy}).
\system then leverages falsification to validate the identified policies on the \comp (Sec.~\ref{subsec:checker}).

\subsubsection{Identifying Physical Channel Policies}
\label{subsec:policy}
To properly designate the circumstances under which the physical interactions are a \emph{vulnerability} or \emph{feature}, we define \emph{intended} and \emph{unintended} labels. 

We define the physical channel between an actuator and an app as intended if the app is installed to be triggered from that command's influence. 
For instance, consider a user that installs an app that turns on the AC when the temperature exceeds a threshold.
If the temperature from the oven triggers this app, the channel between {\small{$\mathtt{oven \mhyphen on}$}} and the app is intended, as the oven causes the temperature to exceed the threshold defined by the user.

We define a physical channel between an actuator and an app as unintended if it is unplanned by a system or undesired by a user.  
For instance, consider a user that installs an app with the goal of unlocking the patio door with the motion from her presence. 
If the motion from the vacuum robot triggers this app, the channel between {\small{$\mathtt{robot \mhyphen vacuum \mhyphen on}$}} and the app is unintended, as unlocking the patio door due to the vacuum robot is not desired by the user. 

\shortsectionBf{Intended/Unintended Label Generation.}
\system automates generating the interaction labels based on the use cases of apps. 
It also allows users to change the labels as their needs dictate. 

To generate the labels, \system first checks whether the intended use of an app is related to a specific activity. 
\system leverages SmartAuth~\cite{tian2017smartauth}, an NLP-based technique that extracts the activity related to an app from its description.

Consider an app that states \textit{``open the windows when you are cooking''} in its description, and the app is triggered when the temperature sensor's readings exceed a threshold.
SmartAuth outputs that this app is related to the cooking activity.
\system takes the app's activity and checks whether any of the actuators installed in the smart home are semantically related to the activity.
For this, \system uses Word2Vec representations to compute the semantic distance between the activity and the commands. It then assigns intended ({\small{$\mathtt{Int}$}}) label to the commands with a distance lower than a threshold and assigns unintended ({\small{$\mathtt{UnInt}$}}) labels to others. 
For instance, the cooking activity is semantically related to {\small{$\mathtt{oven \mhyphen on}$}} and {\small{$\mathtt{cooker\mhyphen on}$}} commands, and thus, \system assigns {\small{$\mathtt{Int}$}} to them.

If an app's description does not indicate an activity or none of the commands are semantically related to the activity, \system assigns the labels based on the apps' sensor events. 
The apps conditioned on motion or sound sensors' events are used to detect the presence of users and intruders in smart homes. For instance, {\small{$\mathtt{App_2}$}} in Sec.~\ref{sec:background}, which unlocks the patio door and turns on the lights when motion is detected, is used to be triggered with user presence. 
\system assigns {\small{$\mathtt{UnInt}$}} label to all commands for such apps because only the influences from users and intruders are intended for them.

\begin{figure}[t!]
    \centering
    \includegraphics[width=\linewidth]{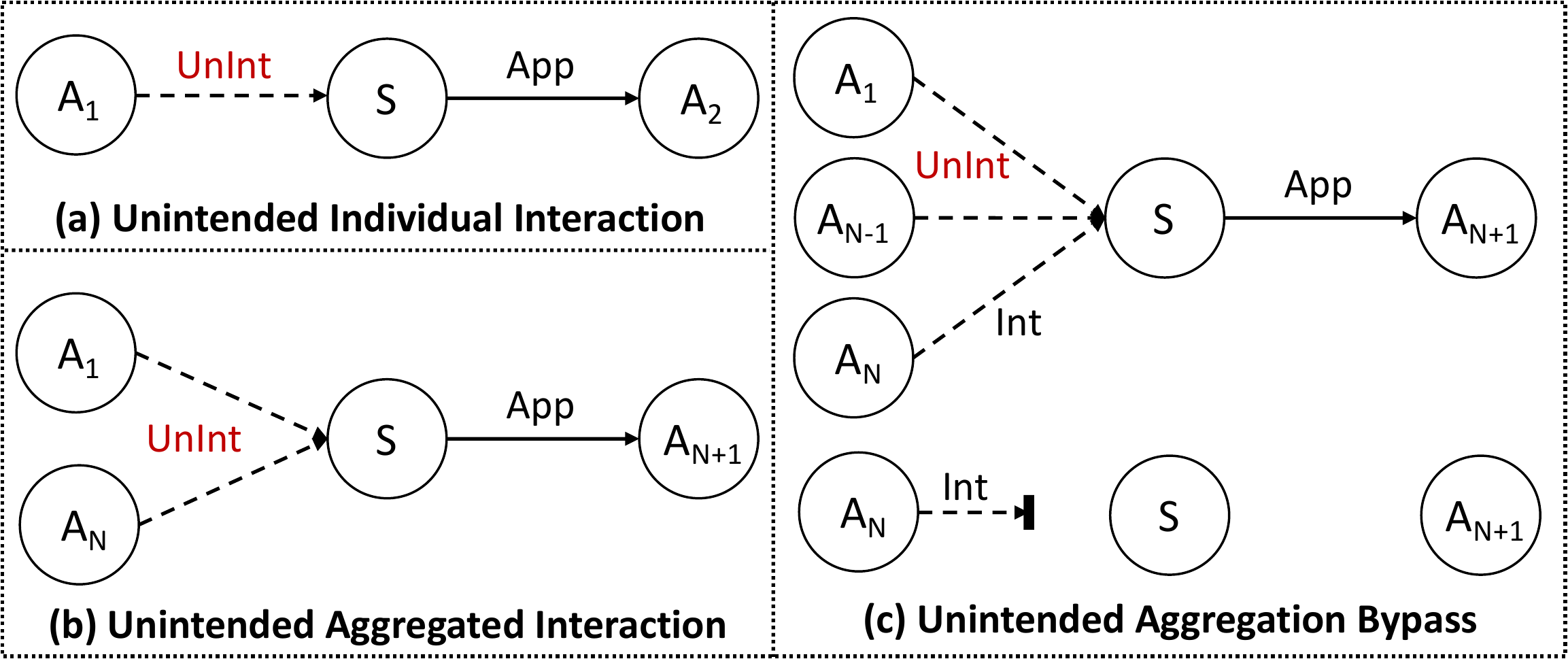}
    \caption{Intent-based physical channel vulnerabilities (A is an actuation command, and S is a sensor event).}
    \label{fig:policy}
\end{figure}

The apps conditioned on temperature, humidity, smoke, and illuminance channels are installed to be triggered when the physical channel's state reaches a specific condition.
For instance, an app that turns on the AC when the temperature is higher than a threshold controls the AC based on the ambient temperature level. 
\system assigns {\small{$\mathtt{Int}$}} label to all commands for such apps because their intended use only depends on the physical channel conditions, regardless of which commands influence them.

\shortsectionBf{Intent-based Policies.}
Based on generated labels, we define three \emph{intent-based} policies, as shown in Figure~\ref{fig:policy}, which are used to identify unintended physical interactions that create undesired and unsafe system states.  
We present the security goal of each policy and its expression with Metric Temporal Logic (MTL) in Table~\ref{tab:policy}. We validate the policies on the \comp in the next section.

The first policy, {\small{$\mathtt{G_1}$}}, unintended individual interaction, states that an actuation command's unintended influence on a physical channel must not trigger an app's sensor event and invoke its device actions (Figure~\ref{fig:policy}(a)).
For instance, the robot vacuum's motion must not trigger an app that unlocks the patio door when a {\small{$\mathtt{motion \mhyphen detected}$}} event occurs. 
This is because an adversary who can invoke the start robot vacuum action (\eg through a vulnerable app) can exploit this interaction to indirectly unlock the patio door.

The second policy, {\small{$\mathtt{G_2}$}}, states the aggregated influence from multiple commands must not unintentionally trigger an app and invoke its device actions (Figure~\ref{fig:policy}(b)).
Although a command's individual influence may not trigger an app, its aggregation with another command's influence may trigger it.
For example, the aggregated sound of {\small{$\mathtt{AC \mhyphen on}$}} and {\small{$\mathtt{dryer \mhyphen on}$}} must not trigger an app that sounds an alarm when the {\small{$\mathtt{sound \mhyphen detected}$}} event occurs, and the home mode is away. 

The last policy, {\small{$\mathtt{G_3}$}}, states if the commands' intended influences do not trigger an app's sensor event, their aggregation with other commands' unintended influences must not trigger it (Figure~\ref{fig:policy}(c)).
For instance, if the light bulb's {\small{$\mathtt{Int}$}} influence on illuminance does not create a {\small{$\mathtt{light \mhyphen detected}$}} event and trigger apps, its aggregation with the TV's {\small{$\mathtt{UnInt}$}} influence must not trigger the apps as well.

These unintended physical interactions, by definition, are not features as they are not desired by users.
Yet, an adversary can exploit them to indirectly control devices and cause unsafe states.

\input{0-CCS2022-FinalVersion/tables/device-centric}

\shortsectionBf{Device-Centric Policies.}
While intent-based policies detect unsafe states from unintended physical interactions, {\small{$\mathtt{Int}$}} labeled physical channels can also cause unsafe states.
For instance, the heater's intended influence on the temperature sensor may trigger an app that opens the windows when the temperature exceeds a threshold.

To address such violations, we extend the security rules of previous works~\cite{celik2018soteria,celik2019iotguard,ding2021iotsafe} (and enhance them with time-constrained temporal operators in MTL) to define device-centric policies.
We present a subset of device-centric policies in Table~\ref{tab:device-policy} and give the complete list in Appendix~\ref{appendix:deviceRules}.
For instance, the {\small{$\mathtt{DC_5}$}} policy states physical interactions must not prevent an alarm from going off in \emph{two secs} after smoke is detected ({\small{$\mathtt{\square(({smoke} > {th}) \rightarrow \Diamond_{[0,2]} ({alarm} = {ON}))}$}}, where {\small{$\mathtt{\square}$}} is always, and {\small{$\mathtt{\Diamond_{[0,2]}}$}} is eventually within next $\mathtt{2}$ secs).

\subsubsection{Validating Policies on \comp}
\label{subsec:checker}
After identified policies are expressed with MTL, \system executes the \comp (hybrid I/O automaton) and collects actuator and sensor traces to validate policies. 

At each execution, the \comp takes apps' activation times as input, which is the time when the app invokes its commands and the command \apem transitions to the ``on'' state.
The \comp simulates the unified physical behavior of commands and sensor events and outputs traces of \apems. 
The traces $\mathtt{(v, t)}$ are composed of a periodic timestamp $\mathtt{t}$, and a physical channel value $\mathtt{v}$.
Each command \apem's $\mathtt{v}$ shows how much it influences a channel, and each sensor event \apem's $\mathtt{v}$ shows its measurements.
The traces also include labels ({\small{$\mathtt{Int/UnInt}$}}) and app IDs of commands/events for root cause analysis.

\shortsectionBf{Policy Validation Challenges.}
The physical channel values and the app activation times are continuous; thus, the \comp's state space becomes infinite, which makes formal verification approaches (\eg model checking) undecidable on the \comp~\cite{alur1995algorithmic,henzinger1998s,plaku2009falsification}.

\input{0-CCS2022-FinalVersion/algorithms/reachability}

To address this issue, we initially implemented a grid-testing approach, a commonly applied method for testing CPS and autonomous vehicle software~\cite{corso2020survey,norden2019efficient,zutshi2014multiple}.
Grid-testing determines whether the \comp satisfies a policy under a finite set of apps' activation times-- the times that apps invoke actuation commands.
Algorithm~\ref{alg:reachability} presents the grid-testing approach on the \comp for policy validation.
We set apps' activation times as a grid ({\small{$\mathtt{t_0}:\Delta \mathtt{t}:\mathtt{t_{end}}$}}) (Line $3$). 
The algorithm executes the \comp with a search on activation time combinations. It then validates a policy on each execution's traces from \apems with a robustness metric (Lines $4\mhyphen 6$), where negative robustness values indicate a policy violation.

However, we found that grid-testing does not scale larger analyses with the increasing number of interacting apps and may miss policy violations due to input discretization.
To address these, we extend optimization-guided falsification and compare it with grid-testing in identifying violations and performance overhead in Sec.~\ref{sec:eval}.

\begin{figure}[t!]
    \centering
    \includegraphics[width=\linewidth]{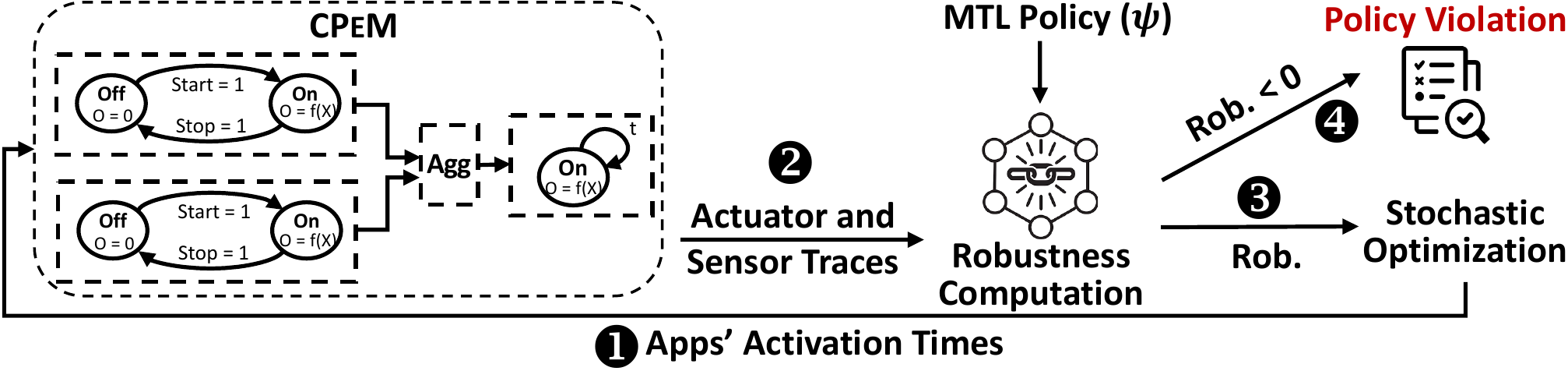}
    \caption{Overview of falsification to find policy violations.}
    \label{fig:falsification}
\end{figure}

\shortsectionBf{Optimization-Guided Falsification.}
Falsification is a formal analysis technique that searches for a counterexample to an MTL policy from a continuous input set~\cite{abbas2013probabilistic,annapureddy2010ant}.
Figure~\ref{fig:falsification} depicts our approach in leveraging falsification to search for interacting apps that cause a policy violation on the \comp.  

Specifically, we use an optimization algorithm to search for policy violations by sampling activation times (\circled{1}). 
We then execute the \comp and record actuator and sensor traces from \apems (\circled{2}).
From the traces, we compute a robustness value that quantifies how close an MTL formula is to the policy violation (\circled{3}). 
Positive robustness values indicate the policy is satisfied, and negative values indicate it is violated.
The sampler then seeds another input to the \comp within the ranges (similar to input mutation in fuzzing~\cite{godefroid2008automated}).
The sampler's objective is minimizing the robustness to find a policy violation (\circled{4}). 
The termination criteria for input generation is when the policy is violated or a user-defined maximum number of iterations is met. 

\begin{lstlisting}[language=json,firstnumber=1,float=ht,
  floatplacement=!ht, caption={An example output of policy validation.}, captionpos=b, belowskip=-1 \baselineskip, aboveskip=-1pt, float=ht, backgroundcolor=\color{light-gray}, basicstyle=\ttfamily, basicstyle=\scriptsize, label = {lst:falsification}]
"Policy Violation": {
  "Input": robot-vacuum-on,
  "Apps":
    "app_a": e:= timer, a:= robot.vacuum-on,
    "app_b": e:= mot-active, a:=patDoor.unlock,
  "Activation Time": 10,
  "Distance": 2,
  "Physical Channel Values": 
    "t = 10": robot-vacuum.move,
    "t = 11": (UnInt) mot-active,
    "t = 11": patDoor.unlock}
\end{lstlisting}

When \system identifies a policy violation, it outputs a quintuple, {\small{$\mathtt{O = (inputs,}$}} {\small{$\mathtt{apps,}$}} {\small{$\mathtt{dist,}$}} {\small{$\mathtt{atime, v)}$}}, that details the policy violation's root cause.
Here, {\small{$\mathtt{atime}$}} is the activation time of {\small{$\mathtt{apps}$}}, {\small{$\mathtt{v}$}} is {\small{$\mathtt{Int, UnInt}$}} labeled command \apem and sensor event \apem outputs, and {\small{$\mathtt{dist}$}} is the distance from actuators to sensors.
Listing~\ref{lst:falsification} presents the output of a policy violation when {\small{$\mathtt{App_a}$}} turns on a robot vacuum and interacts with {\small{$\mathtt{App_b}$}} that unlocks the patio door when it detects the robot vacuum's motion.
The output further details the violation occurs when {\small{$\mathtt{robot \mhyphen vacuum \mhyphen on}$}} is activated at minute $\mathtt{10}$ when it is $\mathtt{2}$ meters away from the motion sensor.
In Sec.~\ref{sec:discussion}, we detail how this output can be used to mitigate the violation.

%% file: 0-CCS2022-FinalVersion/algorithms/composition.tex
\begin{algorithm}[t!]
    \footnotesize
    \setstretch{0.95}
	\caption{Composition of Physical Behavior of Apps}
	\label{alg:composition}
	\begin{algorithmic}[1]
	    \Require{Actuation Command \apems ($\mathtt{H_a}$), Sensor Event \apems ($\mathtt{H_s}$), Apps ($\mathcal{L}_{\mathtt{app}}$)} 
	    \Ensure{\comp ($\mathcal{M}$)} 
	    \vspace{.5mm}
	    \Function{Composition}{$\mathtt{H_a, H_s,} \mathcal{L}_{\mathtt{app}}$}
	   \For{$\mathtt{H_i \in H_a}$, $\mathtt{H_j \in H_s}$}
	   \State \textsc{Unify}$\mathtt{(H_i, H_j)}$ \Comment{\texttt{Command to sensor event transitions}} 
	   \EndFor
	   \For{$\mathtt{app_i \in} \mathcal{L}_{\mathtt{app}}$}
   	        \State $\langle \mathcal{L}^i_{\mathtt{command}}, \mathcal{L}^i_{\mathtt{condition}}, \mathcal{L}^i_{\mathtt{event}} \rangle =$ \textsc{StaticAnalysis(}$\mathtt{app_i}$)
            \For{$\mathtt{s} \in \mathcal{L}^i_{\mathtt{event}}$, $\mathtt{a} \in \mathcal{L}^i_{\mathtt{command}}$}
                \If{$\mathtt{a.condition = true}$}
               \State \textsc{Unify}$\mathtt{(H_s, H_a)}$ \Comment{\texttt{Sensor event to command transitions}}
               \EndIf
   	        \EndFor 
   	    \EndFor
   	    \For{$\mathtt{a_1 \in} \mathcal{L}_{\mathtt{command}}$}
   	    \If{$\mathtt{a_1 \in} \mathcal{L}^i_{\mathtt{event}}$ and $\mathtt{a_i.condition = true}$}
   	    \State \textsc{Unify}$\mathtt{(H_{a_1}, H_{a_i})}$ \Comment{\texttt{Software channel transitions}}
   	    \EndIf
   	    \EndFor
	   \For{$\mathtt{H_j \in H_s}$}
	   \State \textsc{Agg}$\mathtt{(H_j.U)} $ \Comment{\texttt{Aggregate inputs of the sensor events}}
   	   \For{$\mathtt{H_k \in H_s}$}
    	   \If{$\mathtt{H_j.O = H_k.U}$} \textsc{Dep}$\mathtt{(H_j \rightarrow H_k)}$ \Comment{\texttt{Dependency}} 
    	   \EndIf
	    \EndFor
	   \EndFor
	    \State \Return $\mathcal{M} = \bigcup \mathtt{(H_a, H_s)}  $ \Comment{\texttt{Return \comp}}
        \EndFunction
	\end{algorithmic}
	\vspace{-2pt}
\end{algorithm}

%% file: 0-CCS2022-FinalVersion/tables/policies.tex
\begin{table*}[ht!]
\caption{Descriptions of intent-based policies to discover physical channel vulnerabilities in Figure~\ref{fig:policy}.}
    \label{tab:policy}
    \centering
    \setlength{\tabcolsep}{0.4em}
    \def\arraystretch{1.05}
    \resizebox{\textwidth}{!}{
    \begin{threeparttable}
        \begin{tabular}{|c|c|l|l|} 
         \thickhline
         \textbf{ID$^*$} & \textbf{Formal Representation$^\dagger$}  & \multicolumn{1}{c|}{\textbf{Policy Description}}  &  \multicolumn{1}{c|}{\textbf{Security Goal}} \\ \thickhline
         \multirow{2}{*}{$\mathtt{G_1}$}  &  \multirow{2}{*}{$\square (\mathtt{imp(\langle UnInt,p \rangle) \le th}) $} & An actuation command's influence on a physical channel must not & Prevent attackers from creating and exploiting \\ 
         & & unintentionally trigger an app's sensor event and invoke its device actions. & individual unintended physical interactions. \\ \hline

         \multirow{2}{*}{$\mathtt{G_2}$} & \multirow{2}{*}{$\mathtt{\square ({imp(\langle UnInt_1,p \rangle, \ldots, \langle UnInt_n,p \rangle)} \le {th}) }$} & Multiple commands' aggregated influence on a physical channel must not  & Prevent attackers from creating and exploiting \\ 
          & & unintentionally trigger an app's sensor event and invoke its device actions. & aggregated unintended physical interactions. \\ \hline

         \multirow{2}{*}{$\mathtt{G_3}$} & $\mathtt{\square  ({imp(\langle Int_1, p \rangle \ldots \langle Int_k, p \rangle)}<{th}) \rightarrow \square  ({imp(\langle Int_1, p \rangle,} }$ & If commands' intended influences do not trigger an app's sensor event, their & Prevent attackers from bypassing intended \\
         & $\mathtt{\ldots, \langle Int_k, p \rangle, \langle UnInt_1,p \rangle, \ldots, \langle UnInt_n,p \rangle) < th}) $ & aggregation with other commands' unintended influences must not trigger it. &  physical interactions.  \\  \hline
        \end{tabular}
          $*$ The IDs correspond to the vulnerabilities in Figure~\ref{fig:policy} ($\mathtt{G_1}$ for (a), $\mathtt{G_2}$ for (b), and $\mathtt{G_3}$ for (c)) 

          $\dagger$ $\mathtt{imp()}$ denotes an actuation commands' labeled influence on a channel $\mathtt{p}$. Multiple channels in $\mathtt{imp}$ denotes aggregated influences. $\mathtt{th}$ denotes a sensor's sensitivity. 
    \end{threeparttable}
    }
\end{table*}

%% file: 0-CCS2022-FinalVersion/tables/device-centric.tex
\begin{table}[t!]
\caption{Example device-centric policies (The complete list of policies is presented in Appendix~\ref{appendix:deviceRules}).} 
    \label{tab:device-policy}
    \centering
    \setlength{\tabcolsep}{0.3em}
    \def\arraystretch{1.05}
    \resizebox{\columnwidth}{!}{
    \begin{threeparttable}
        \begin{tabular}{|c|c|c|} 
         \thickhline
         \textbf{ID} & \textbf{Policy Description} & \textbf{Formal Representation}  \\ \thickhline
         \multirow{2}{*}{$\mathtt{DC_2}$} & When the home is in the away mode, & \multirow{2}{*}{$\mathtt{\square(mode \mhyphen away \rightarrow window \mhyphen close)}$}  \\ 
         & the window must be closed. & \\ \hline
         \multirow{2}{*}{$\mathtt{DC_3}$} & A device must not open, then close and  &  \multirow{2}{*}{$\mathtt{\square\neg(on\wedge\Circle\Diamond_{[0,{t}]}( off\wedge \Circle\Diamond_{[0,{t}]} on))}$} \\
         & then reopen (actuation loop) within $\mathtt{t}$ seconds. &   \\  \hline

        \multirow{2}{*}{$\mathtt{DC_5}$} & The alarm must go off within $\mathtt{t}$ seconds  & \multirow{2}{*}{$\mathtt{\square(smoke \mhyphen detected \rightarrow\Diamond_{[0,{t}]}alarm \mhyphen on)}$}  \\
         &after smoke is detected. & \\\hline
        
        \multirow{2}{*}{$\mathtt{DC_6}$} & The main door must not be left  &   \multirow{2}{*}{$\mathtt{\Diamond_{[0,{t}]}door \mhyphen lock}$} \\
         & unlocked for more than $\mathtt{t}$ seconds. & \\ \hline 
         
         \multirow{2}{*}{$\mathtt{DC_{9}}$} & The door must always be locked and lights  & \multirow{2}{*}{$\mathtt{\square(mode\mhyphen away \rightarrow door \mhyphen lock \wedge light \mhyphen off)}$}  \\ 
         & must be off when the home is in the away mode. & \\ \thickhline
 
        \end{tabular}
              \end{threeparttable}}
\end{table}

%% file: 0-CCS2022-FinalVersion/algorithms/reachability.tex
\begin{algorithm}[t!] 
    \footnotesize
	\caption{Grid-Testing} 
	\label{alg:reachability}
	\begin{algorithmic}[1]
        \Require{\comp ($\mathcal{M}_{\mathtt{H_a, H_s}}$) with command \apems ($\mathtt{H_a}$) and sensor event \apems ($\mathtt{H_s}$), parameters ($\mathtt{x}$ - distances among devices), inputs ($\mathtt{U}$ - apps' activation times $\mathtt{t_0}:\Delta \mathtt{t}:\mathtt{t_{end}}$), policy ($\psi$).}
		\Ensure{$\mathtt{P}$=$\mathtt{(inputs, apps, dist, atime, y)}$}
		\vspace{1mm}
		\Function{Grid\_Test}{$\mathtt{H_a, H_s, x,U}, \mathcal{M}_{\mathtt{H_a, H_s}},\psi$}
		\For{$\mathtt{j} \in \mathtt{H_s}$, $\mathtt{H_{OP}}\subseteq \mathtt{H_a}$} 
		\For{Different activation times in $\mathtt{H_{OP}}$}
		\If{$\Phi(\mathcal{M}_{\mathtt{H_{OP}, H_s}},\mathtt{x},\mathtt{u}) \nvDash \psi$}
		\State $\mathtt{P} \leftarrow \mathtt{P} \cup \{\mathtt{x,u},\Phi(\mathcal{M}_{\mathtt{H_{OP}, H_s}},\mathtt{x},\mathtt{u})\}$
		\EndIf
		\EndFor
		\EndFor
		\State \Return {$\mathtt{P}$}
        \EndFunction
	\end{algorithmic}
\end{algorithm}

%% file: 0-CCS2022-FinalVersion/text/eval-new.tex
\section{Evaluation}
\label{sec:eval}

\begin{figure}[t!]
    \centering
    \includegraphics[width=0.9\linewidth]{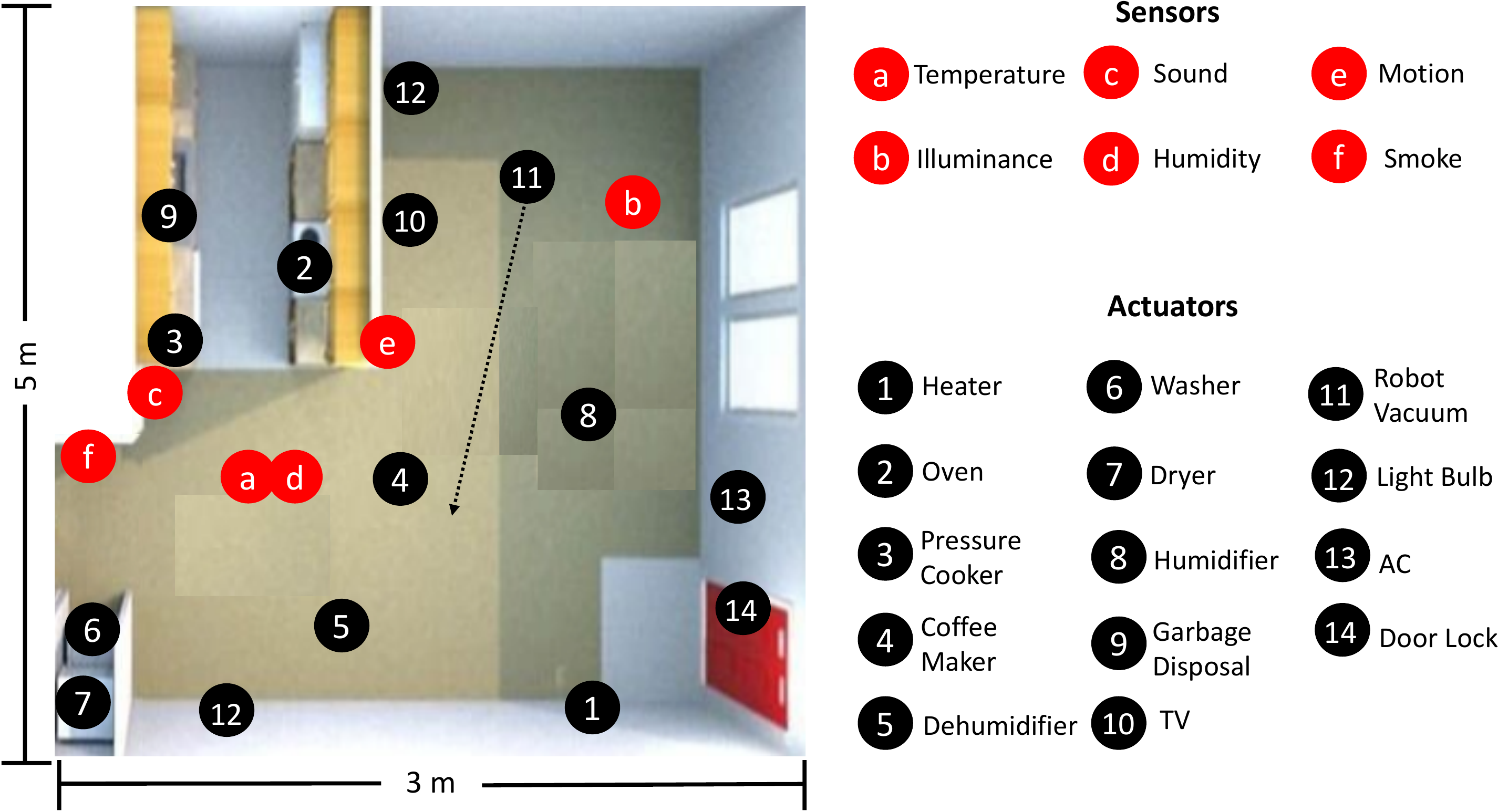}
    \caption{Sensor/actuator layout in the actual house.}
    \label{fig:house}
\end{figure}

We evaluate \system in a real home with six sensors and $14$ actuators, as shown in Figure~\ref{fig:house}.
To automate the devices, we study three IoT app markets, IFTTT, Microsoft Flow, and SmartThings, and install $39$ popular apps (See Appendix~\ref{sec:apps_experiments} for their descriptions).
We then invoke each actuation command and collect sensor measurements from actual devices in the house for $30$ mins to identify the channels each command influences (See Table~\ref{tab:impact})\footnote{We have consulted our university’s IRB office and got advised that IRB approval is not required since we do not collect any sensitive information.}.

We built a total of $24$ \apems for channels the commands influence (number of physical channels the actuators influence in Table~\ref{tab:impact}) and six sensor event \apems for the channels they observe (temperature, illuminance, sound, humidity, motion, and smoke).
We also use the collected sensor measurements to conduct SI and tune the \comp parameters.
We detail the \comp fidelity in Appendix~\ref{appendix:fidelity}.
Lastly, we assign {\small{$\mathtt{Int}$/$\mathtt{UnInt}$}} labels between the commands and apps based on the apps' intended uses, as described in Sec.~\ref{subsec:policy}.

We implement grid-testing and our falsification approach using an open-source temporal logic toolbox, {\small{$\mathtt{\staliro}$}}~\cite{annpureddy2011s}. 
We leverage {\small{$\mathtt{\staliro}$}}'s {\small{$\mathtt{dptaliro}$}} function as a subroutine in grid-testing and {\small{$\mathtt{staliro}$}} function in our optimization-guided falsification to validate MTL policies on the \comp.
We use the {\small{$\mathtt{staliro}$}} function with simulated annealing by hit-and-run Monte Carlo sampling for input generation~\cite{fainekos2019robustness}.
We use a dynamic programming-based algorithm to compute the robustness of MTL policies.
In grid-testing's implementation, we consider apps invoke actuation commands at $\mathtt{10}$-min intervals; thus, the apps' activation times are set as a grid to $\mathtt{0:10:60}$.
In falsification's implementation, we set the apps' activation times as continuous ranges, any time in the execution ($\mathtt{0 \mhyphen 60}$), and we define the max number of tests to $\mathtt{100}$ as we do not observe a significant change in robustness after $\mathtt{100}$.

We run the \comp executions on a laptop with a $2.3$ GHz $2$-core i5 processor and $8$ GB RAM, using Simulink 10.0. 

\input{0-CCS2022-FinalVersion/tables/actuator_impacts}

\input{0-CCS2022-FinalVersion/tables/violations2}

\subsection{Effectiveness}
\label{subsec:effectiveness}

We validate intent-based and device-centric security policies on the \comp tuned for our home (See Appendix~\ref{appendix:simulink} for an example \comp). 

Table~\ref{tab:violation} presents $16$ identified policy violations caused by physical interactions among seven different groups of devices.
We compare the violations flagged by \system with prior works that identify physical interaction vulnerabilities and show they can discover $2$ out of $16$ violations. 
We also conduct in-home experiments with real devices and confirm that all $16$ violations are true positives.

\subsubsection{Intent-based Policy Violations} \system identified $14$ interactions that subvert the apps' intended use, causing unsafe states. 

\shortsectionBf{Individual Policy ($\mathtt{G_1}$) Violations.}
\system flagged $10$ individual policy ($\mathtt{G_1}$) violations that occur due to the physical interactions among three groups of devices. 
First, the motion sensor detects the presence of the robot vacuum and unintentionally triggers five apps conditioned on the {\small{$\mathtt{motion \mhyphen detected}$}} event.
For instance, this violation occurs when {\small{$\mathtt{App_{17}}$}} starts the robot vacuum when the home mode is set to away, and {\small{$\mathtt{App_{27}}$}} unlocks the door when motion is detected.
Second, the sound sensor detects the garbage disposal's ($\mathtt{V_2}$) and TV's ($\mathtt{V_3}$) sound, unintentionally triggering apps conditioned on {\small{$\mathtt{sound \mhyphen detected}$}}.
For example, {\small{$\mathtt{App_{9}}$}} turns on the garbage disposal at a user-defined time, and {\small{$\mathtt{App_{19}}$}} notifies the user when sound is detected, creating unnecessary panic.

\shortsectionBf{Aggregation Policy ($\mathtt{G_2}$) Violations.}
\system flagged three aggregation policy ($\mathtt{G_2}$) violations among one group of devices.
The sound sensor outputs a {\small{$\mathtt{sound \mhyphen detected}$}} event due to the unintended aggregated influence from {\small{$\mathtt{AC\mhyphen on}$}}, {\small{$\mathtt{washer\mhyphen on}$}}, and {\small{$\mathtt{dryer \mhyphen on}$}}. 
This, in turn, triggers three apps conditioned on the {\small{$\mathtt{sound \mhyphen detected}$}} event and causes {\small{$\mathtt{light\mhyphen on}$}}, {\small{$\mathtt{TV \mhyphen on}$}}, and {\small{$\mathtt{call\mhyphen user}$}} actions.

\shortsectionBf{Bypass Policy ($\mathtt{G_3}$) Violations.}
\system identified a single bypass policy ($\mathtt{G_3}$) violation. 
The illuminance sensor measures the aggregated illuminance of {\small{$\mathtt{light\mhyphen on}$}} and {\small{$\mathtt{TV \mhyphen on}$}}. The increase in illuminance triggers {\small{$\mathtt{App_{34}}$}}, turning off the lights.
However, {\small{$\mathtt{App_{34}}$}}'s intended operation is turning off the lights when the daylight is enough to illuminate the environment, which is semantically related to the {\small{$\mathtt{light\mhyphen on}$}} action. Therefore, {\small{$\mathtt{light\mhyphen on}$}}'s influence on {\small{$\mathtt{App_{34}}$}}'s sensor event is intended, whereas {\small{$\mathtt{TV \mhyphen on}$}}'s influence is unintended.
Since {\small{$\mathtt{light\mhyphen on}$}}'s individual influence cannot trigger {\small{$\mathtt{App_{34}}$}}'s {\small{$\mathtt{light\mhyphen detected}$}} event but its influence aggregated with the unintended influence from {\small{$\mathtt{TV \mhyphen on}$}} triggers it, the app's intended use is bypassed.

\subsubsection{Device-Centric Policy Violations}
\system identified two device-centric policy violations, {\small{$\mathtt{V_{6}}$}} between three apps and {\small{$\mathtt{V_{7}}$}} between four apps.
In {\small{$\mathtt{V_{6}}$}}, while the home is in sleep mode, the AC's influence on temperature triggers {\small{$\mathtt{App_{23}}$}} to turn on the bulb. This violates {\small{$\mathtt{DC_{8}}$}} since the bulb is turned on when the home mode is sleep.
In {\small{$\mathtt{V_{7}}$}}, two physical interactions occur due to {\small{$\mathtt{UnInt}$}} and {\small{$\mathtt{Int}$}} channels. 
While the home mode is away, the vacuum's {\small{$\mathtt{UnInt}$}} motion triggers {\small{$\mathtt{App_{26}}$}} to turn on the heater (\circled{\small{1}}). The heater's {\small{$\mathtt{Int}$}} temperature then triggers {\small{$\mathtt{App_{23}}$}} and turns on the bulb, violating {\small{$\mathtt{DC_{9}}$}}.

\subsubsection{Comparison with Previous Work}
In Table~\ref{tab:violation}, we compare the policy violations flagged by \system with the most applicable approaches, iRuler~\cite{wang2019charting}, IoTMon~\cite{ding2018safety}, and IoTSafe~\cite{ding2021iotsafe}, that run on IoT app source code to identify physical interaction vulnerabilities.

To identify physical interactions among apps, iRuler uses device behavioral models (\eg {\small{$\mathtt{AC\mhyphen on}$}} decreases the temperature by {\small{$\mathtt{1 \degree C}$}} every hour), and IoTMon mines the apps' text descriptions (\eg finds AC is semantically related to temperature).
If we assume they correctly map all physical channels that each command influences in our smart home, iRuler cannot identify any of \system's violations, and IoTMon can identify $2$ out of $16$ violations.
This is because ($1$) their policies cannot reason about the intended use of apps, ($2$) they do not consider the complex physical properties such as aggregation and dependency, and ($3$) iRuler does not consider device-centric vulnerabilities. 
Additionally, IoTMon would flag $18$ false positives as most commands do not individually cause physical interactions.
To illustrate, it defines a physical channel between the temperature sensor and oven; yet {\small{$\mathtt{oven \mhyphen on}$}}'s individual influence on temperature is not enough to cause an interaction.

IoTSafe models the apps' physical behavior through dynamically collected sensor traces and predicts physical channel values at run-time for policy enforcement.
Compared to IoTMon, IoTSafe does not flag any false positives since it relies on sensor traces collected from real devices.
However, it can also detect only $2$ out of $16$ violations in our smart home.
This is because, as a run-time enforcement system, it cannot infer the influence of an exact command on a physical channel. 
Additionally, its policy enforcement may create unnecessary panic in our home since the robot vacuum's motion would trigger its policy that sounds an alarm and sends a message to the user when motion is detected.

\subsubsection{In-Home Validation Experiments}
\label{subsec:inhouse}
We repeated each identified violation in the actual house and confirmed that they ({\small{$\mathtt{V_1}$-$\mathtt{V_{7}}$}}) are true positives.
In-home validation begins by analyzing each violation's root cause through their ({\small{$\mathtt{inputs,}$}} {\small{$\mathtt{apps,}$}} {\small{$\mathtt{dist,}$}} {\small{$\mathtt{atime, y}$}}) logged by \system.
We activate each actuator involved in the violation using its {\small{$\mathtt{inputs}$}} and {\small{$\mathtt{atime}$}}.
We record the traces and confirm the interacting {\small{$\mathtt{apps}$}}. 
Lastly, we compare traces from \apems and devices to identify any differences in the time the violations occur.  

We observe that policy violations on continuous physical channels (temperature) occur later in the house compared to \comp.
For instance, when {\small{$\mathtt{heater \mhyphen on}$}} is invoked at time $\mathtt{0}$ in the \comp, \system flags a {\small{$\mathtt{DC_9}$}} violation at second $\mathtt{56}$. However, we observe the violation at second $\mathtt{\approx 65}$ in the house.
In contrast, violations on instant channels (motion, sound, and illuminance) occur with minor time deviations.
These slight timing deviations are expected due to inevitable environmental noise.

\begin{figure}[t!]
    \centering
    \includegraphics[width=\linewidth]{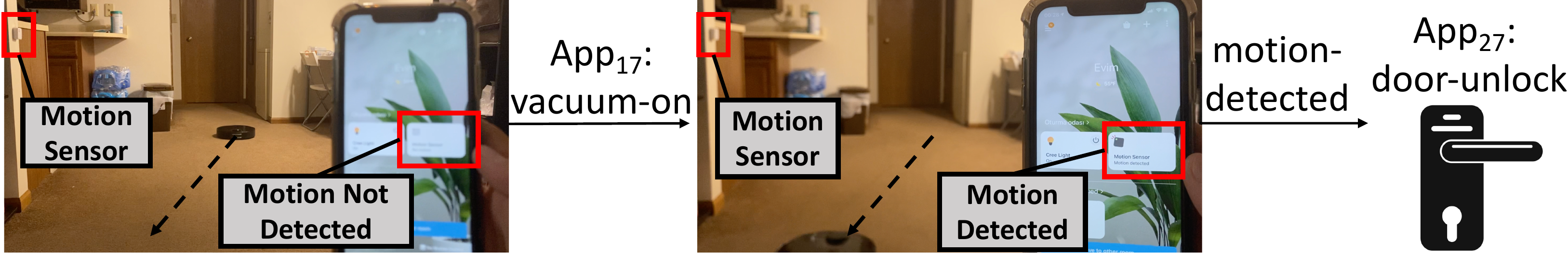}
    \caption{Illustration of the $\mathtt{V_1}$ violation.}
    \label{fig:v1}
\end{figure}

\shortsectionBf{Case Study.} 
We present a case study to illustrate a violation and detail an attack scenario demonstrating how an adversary can exploit the observed physical interaction.
Figure~\ref{fig:v1} depicts the {\small{$\mathtt{V_1}$}} violation; a {\small{$\mathtt{motion \mhyphen detected}$}} event occurs due to the {\small{$\mathtt{robot \mhyphen vacuum \mhyphen start}$}} command and triggers {\small{$\mathtt{App_{27}}$}} that unlocks the door.
To exploit this, an adversary can leverage a vulnerability in the vacuum controller app to start it and stealthily unlock the door. 
The adversary can also wait until the user sets the home mode to away, which triggers {\small{$\mathtt{App_{17}}$}} that turns on the vacuum and causes the door to unlock.
Through this attack, the adversary can break into the house.

\subsection{Violations with Device Placement Changes} 
We change the placement of illuminance, sound sensors, and the TV and use \system to identify the policy violations that occur with the new placement. 
We select these devices as they are easily relocated and potentially impact nine policy violations.

\system correctly identifies that three physical channels that caused policy violations in the initial placement do not occur anymore, and it discovers two new physical channels causing violations.
We then evaluate the impact of the distance parameter's accuracy in identifying policy violations and show that \system only misses a single violation when the distance parameter has a $50\%$ error.

\shortsectionBf{Violations After New Device Placement.} Table~\ref{tab:placement_eval} presents the physical channels that cause policy violations after the device placement changes.
\system identified three physical channels ({\small{$\mathtt{V_{2}}$}}, {\small{$\mathtt{V_{4}}$}}, and {\small{$\mathtt{V_{5}}$}}) that caused policy violations before do not occur in the new device placement. On the contrary, {\small{$\mathtt{TV \mhyphen on}$}} command still creates a {\small{$\mathtt{sound \mhyphen detected}$}} event, unintentionally triggering two apps.

\system also flagged two new physical channels that cause policy violations, where the washer's and dryer's sound cause {\small{$\mathtt{G_{1}}$}} violations since the sound sensor was moved closer to them.
We confirmed with in-home experiments that all identified policy violations with the new device placement are true positives.

\input{0-CCS2022-FinalVersion/tables/placement_eval}

\shortsectionBf{\system's Tolerance to Errors in Distance Parameter.}
Errors in the distance parameter may occur due to slight deviations in the accuracy of IoT device localization tools and RSSI-based localization techniques.
Such errors do not impact \system's ability to discover the initial policy violations since \system integrates SI to tune the \apems based on real device traces.
However, the errors may impact \system's effectiveness in identifying violations after a device's location is changed.
Thus, we introduce errors ranging from {\small{$\mathtt{\pm 10\%}\mhyphen$}}{\small{$\mathtt{50\%}$}} to the distance parameter when the placements of devices are changed and check if \system outputs any false positive or negative violations.
We select these error rates because \system's RSSI-based distance estimation gives, on average, a $22.8\%$ error in our setup, and the only error higher than $50\%$ is the estimation between the TV and sound sensor (due to the wall between devices).

Table~\ref{tab:placement_eval} presents the maximum error in distance parameters under which \system can still correctly identify the violations.
For instance, if \system estimates the distance between the washer and sound sensor with a $20\%$ error, it can still correctly identify these devices' policy violations. However, if the error is larger than $20\%$, \system would miss the violations, causing false negatives.
On the contrary, \system's dryer and TV \apems are more tolerant to error, where they correctly identify the violations even under $50\%$ error. 
This is because these devices' influences on sound are higher, enabling \system to identify their interactions. 

We further checked whether the errors in the distance parameter cause any false positives, where \system flags violations that do not actually occur. 
We found that \system identifies two false positives if the distance error is $50\%$, ($1$) the {\small{$\mathtt{bulb \mhyphen on}$}}'s illuminance \apem, and ($2$) the {\small{$\mathtt{AC \mhyphen on}$}}'s sound \apem.
Yet, such high errors are unlikely in practice since \system integrates state-of-the-art localization techniques.

\subsection{Performance Evaluation}
\label{subsec:perf}

\subsubsection{Scalability Experiments}
We evaluate the policy validation time of \system's falsification algorithm and compare its results with the baseline approach of grid-testing.
Grid-testing discretizes the times when apps activate actuators as a grid and validates policies with all combinations of activation times. 
Although grid-testing identifies all $16$ violations in our experiments, it yields a high time overhead, as detailed below.

\begin{figure}[t]
\begin{subfigure}{.23\textwidth}
  \centering
  \includegraphics[width=\linewidth]{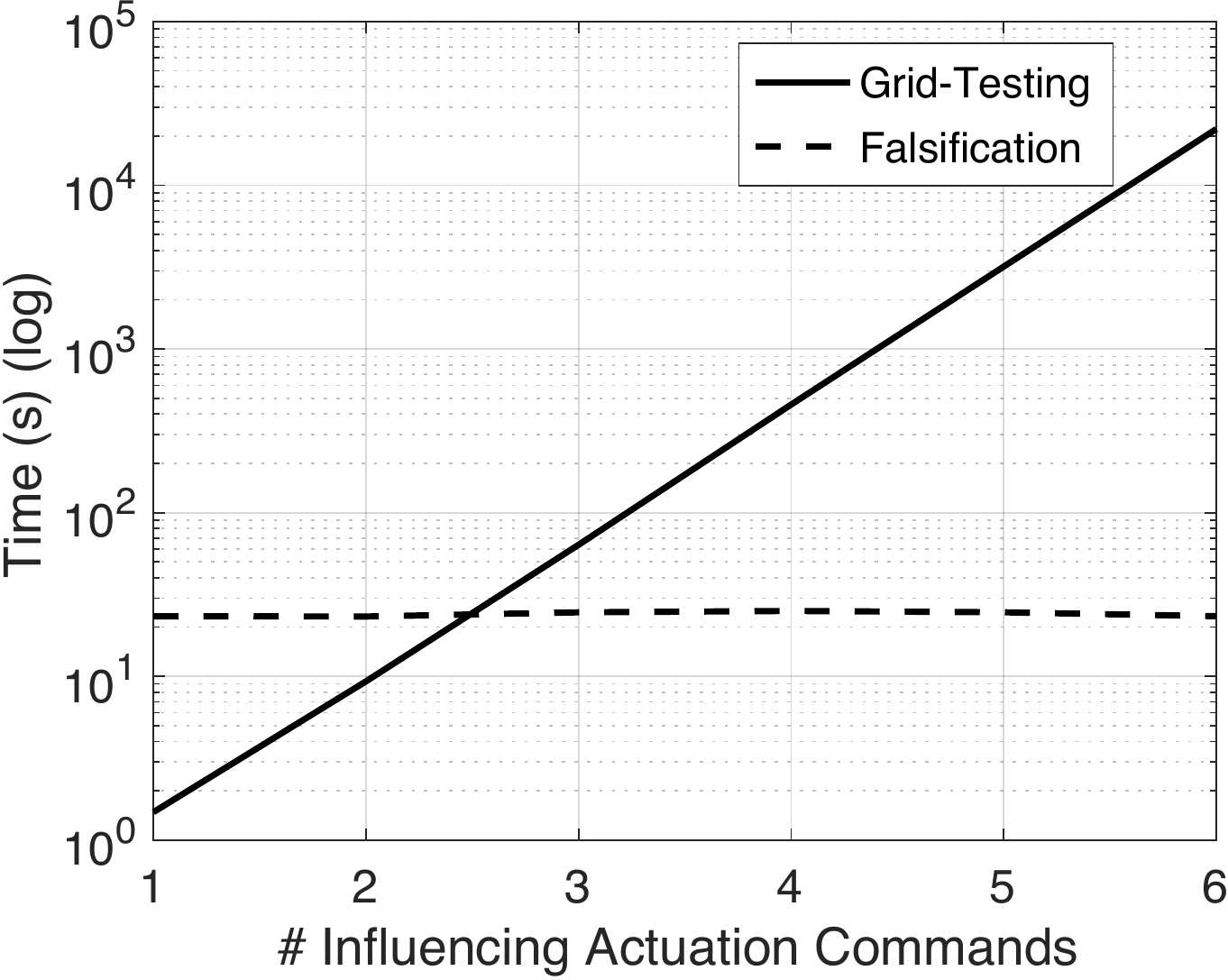} 
  \caption{}
  \label{fig:actuator_time}
\end{subfigure}
\begin{subfigure}{.24\textwidth}
  \centering
  \includegraphics[width=\linewidth]{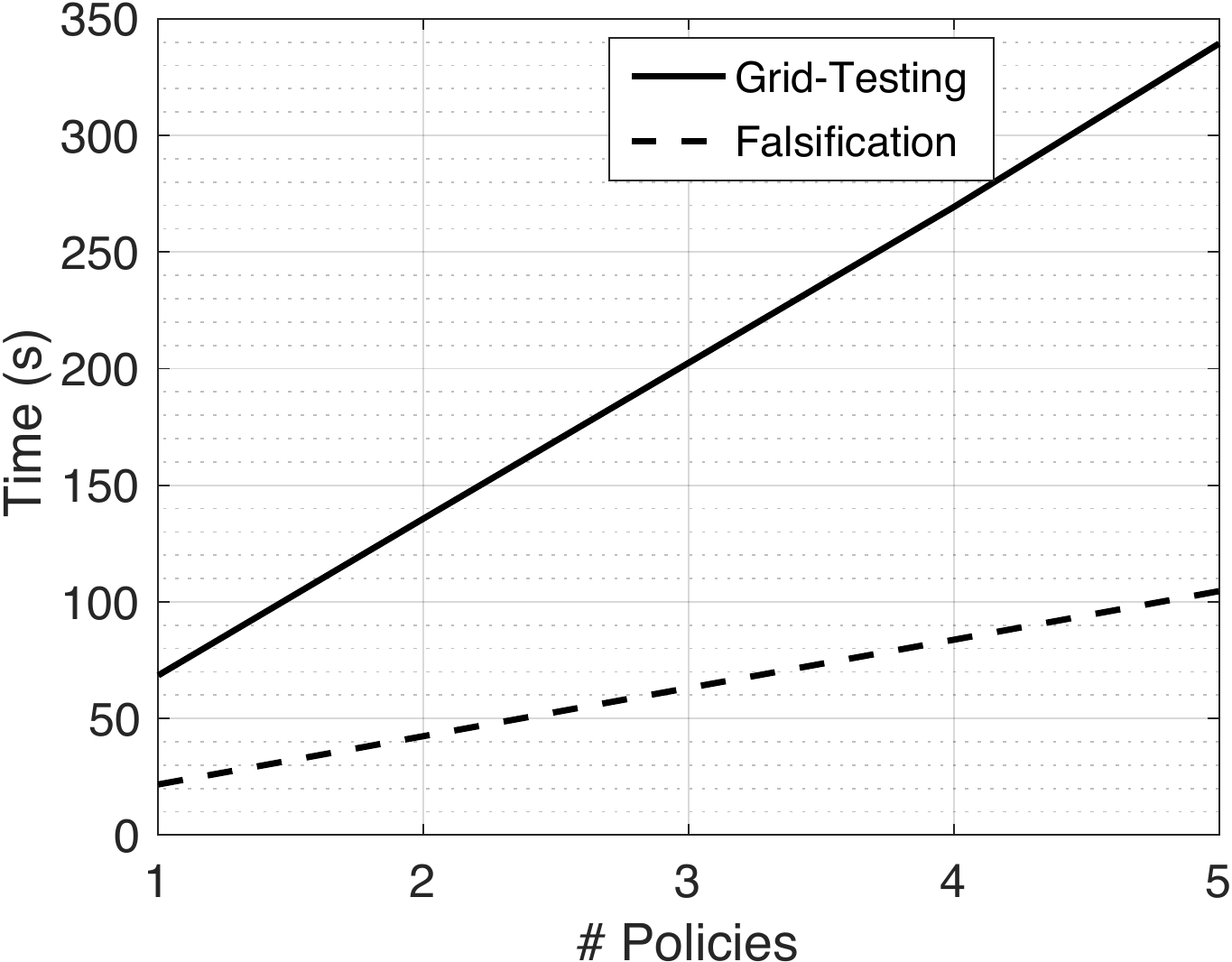}
  \caption{}
  \label{fig:policy_time}
\end{subfigure}
\caption{{(a) \#Actuation commands and (b) \#Policies vs. time.}}
\label{fig:perf}
\end{figure}

\shortsectionBf{\comp Size vs. Time.} 
Figure~\ref{fig:actuator_time} shows the policy validation time of the grid-testing and falsification with an increasing number of command \apems influencing the same channel in the \comp. %
Testing time exponentially increases with the number of commands as each policy is validated using a combination of apps' activation times.
In contrast, falsification has a near-constant time overhead as it samples activation times and searches for low robustness values for a single violation. This adds a negligible delay, on the order of seconds, with an increasing number of commands.

\shortsectionBf{Number of Policies vs. Time.} 
We evaluate the validation time with an increasing number of policies. 
We set the number of command \apems to three, and a sensor measures their aggregated physical influence.
Figure~\ref{fig:policy_time} shows the time overhead of both testing and falsification increases linearly with the number of policies. 
Here, falsification is {\small{$\mathtt{\approx 3\times}$}} more efficient than testing as testing validates the policies with all combinations of apps' activation times.

\subsubsection{Time for Device Trace Collection}
We present the time for actuator and sensor trace collection from actual devices to tune \comp parameters.
It took $\mathtt{7}$ hours to record the measurements from actual sensors, which required turning on each actuator and collecting traces from all sensors in the smart home.
This is an improvement over generating generic flow functions by SI solely using device traces, which requires $\mathtt{\approx 175}$ hours of data collection with different device properties and distances.

%% file: 0-CCS2022-FinalVersion/tables/actuator_impacts.tex
\begin{table}[t!]
    \centering
     \caption{Physical channels of studied actuators and sensors.}
    \label{tab:impact}
    \setlength{\tabcolsep}{0.42em}
    \def\arraystretch{0.9}
    \resizebox{\columnwidth}{!}{
    \begin{threeparttable}
        \begin{tabular}{c|c|c|c|c|c|c|} 
         \cline{2-7}
         &  \multicolumn{6}{c|}{\textbf{Sensors}}  \\ \hline
         \multicolumn{1}{|c|}{\textbf{Actuator (Actuation Command)}} &  \textbf{Temp.} & \textbf{Illum.} & \textbf{Sound} & \textbf{Hum.} & \textbf{Motion} & \textbf{Smoke} \\ \hline
          \multicolumn{1}{|c|}{Marley Baseboard Heater (set(val))}& \cmark & \xmark & \xmark & \cmark & \xmark & \xmark \\ \hline

          \multicolumn{1}{|c|}{Kenmore AC (set(val))}& \cmark & \xmark & \cmark & \cmark & \xmark & \xmark \\ \hline

          \multicolumn{1}{|c|}{CREE Smart Light Bulb (on)} & \xmark  & \cmark & \xmark & \xmark &\xmark & \xmark\\ \hline

          \multicolumn{1}{|c|}{Instant Pot Pressure Cooker (on)} & \cmark & \xmark & \xmark & \cmark &\xmark & \xmark \\ \hline

          \multicolumn{1}{|c|}{Mr. Coffee Coffee maker (on)} & \cmark & \xmark & \xmark & \cmark &\xmark & \xmark \\ \hline

         \multicolumn{1}{|c|}{Sunbeam Humidifier (on)} & \xmark  & \xmark & \xmark & \cmark &\xmark & \xmark\\ \hline

         \multicolumn{1}{|c|}{Easy Home Dehumidifier (on)} & \xmark  & \xmark & \xmark & \cmark &\xmark & \xmark\\ \hline

          \multicolumn{1}{|c|}{Whirlpool Clothes Washer (on)} & \xmark & \xmark & \cmark & \cmark & \xmark &\xmark \\ \hline
          
          \multicolumn{1}{|c|}{Whirlpool Dryer (on)} & \cmark & \xmark & \cmark & \cmark & \xmark & \xmark \\ \hline

           \multicolumn{1}{|c|}{Whirlpool Garbage Disposal (on)} & \xmark & \xmark & \cmark & \xmark & \xmark & \xmark \\ \hline

          \multicolumn{1}{|c|}{Roborock S4 Robot Vacuum (on)} & \xmark & \xmark & \xmark & \xmark & \cmark & \xmark \\ \hline

        \multicolumn{1}{|c|}{Vizio 48'' TV (on)} & \xmark & \cmark & \cmark & \xmark & \xmark & \xmark \\ \hline
        
        \multicolumn{1}{|c|}{Door (unlock)}& \xmark & \xmark & \cmark & \xmark & \xmark & \xmark \\ \hline
        
        \multicolumn{1}{|c|}{Whirlpool Oven (on)} & \cmark & \xmark & \xmark & \cmark & \xmark & \xmark \\ \hline
        
        \end{tabular}
          \xmark~means the command does not influence the physical channel the sensor observes and \cmark means the command influences it. See Appendix~\ref{appendix:details} for sensor brands.
    \end{threeparttable}
    }
\end{table}

%% file: 0-CCS2022-FinalVersion/tables/violations2.tex
\begin{table*}[t!]
    \centering
    \def\arraystretch{1.15}
    \caption{Policy violations identified by \system and previous works.}
    \label{tab:violation}
    \resizebox{\textwidth}{!}{
    \begin{threeparttable}
        \begin{tabular}{|c|c|l|c|c|c|c|c|} 
        \thickhline
         \multirow{2}{*}{\textbf{Policy}} & \multirow{2}{*}{\textbf{ID}} & \multicolumn{1}{c|}{\multirow{2}{*}{\textbf{App Interactions}}} & \multirow{1}{*}{\textbf{Number of}} & \multirow{2}{*}{\textbf{Violation Description}} & \multicolumn{3}{|c|}{\textbf{Existing Work}} \\ \cline{6-8}
         
         &  & & \multirow{1}{*}{\textbf{Violations}} &  & \textbf{iRuler} & \textbf{IoTMon} & \textbf{IoTSafe} \\ \thickhline

        \multirow{9}{*}{$\mathtt{G_1}$} &\multirow{5}{*}{$\mathtt{V_1}$}& \multirow{1}{*}{$\hspace{28mm}\mathtt{\myrightarrow{\mathtt{motion\mhyphen det.}} door \mhyphen unlock}$}  & \multirow{5}{*}{$5$} & \multirow{3}{*}{Robot vacuum's motion \textit{unintentionally} triggers} & \multirow{5}{*}{\xmark} & \multirow{5}{*}{\xmark} & \multirow{5}{*}{\xmark}  \\ 
        & &  \multirow{1}{*}{$\mathtt{\hspace{28mm}\myrightarrow{\mathtt{motion\mhyphen det.}} light \mhyphen on}$}  & & \multirow{3}{*}{five apps and causes door-unlock, heater-on,} & & & \\
        & &  \multirow{1}{*}{$\mathtt{robot\mhyphen vacuum \mhyphen start \myrightarrow{\mathtt{motion\mhyphen det.}} heater \mhyphen on}$}  &  & \multirow{3}{*}{light-on, TV-on and call-user actions.} & & & \\ 
        & &  \multirow{1}{*}{$\hspace{28mm}\mathtt{\myrightarrow{\mathtt{motion\mhyphen det.}} TV \mhyphen on}$}  &  &  & & & \\ 
        & &  \multirow{1}{*}{$\mathtt{\hspace{28mm}\myrightarrow{\mathtt{motion\mhyphen det.}} call \mhyphen user}$}  & & & & & \\ \cline{2-8}  
        
         & \multirow{3}{*}{$\mathtt{V_2}$} & \multirow{1}{*}{$\mathtt{\hspace{19mm}\myrightarrow{\mathtt{sound\mhyphen det.}} light \mhyphen on}$}  & \multirow{3}{*}{$3$}  & Garbage disposal's sound \textit{unintentionally} & \multirow{3}{*}{\xmark} & \multirow{3}{*}{\xmark} & \multirow{3}{*}{\xmark} \\
         & & \multirow{1}{*}{$\mathtt{garb\mhyphen disp \mhyphen on \myrightarrow{\mathtt{sound\mhyphen det.}} TV \mhyphen on}$} &  & triggers three apps and causes  & & & \\ 
          & & \multirow{1}{*}{$\mathtt{\hspace{19mm}\myrightarrow{\mathtt{sound\mhyphen det.}} call \mhyphen user}$} &  & light-on, TV-on and call-user actions. & & & \\ \cline{2-8}
          
          & \multirow{2}{*}{$\mathtt{V_3}$} & \begin{tabular}[t]{@{}c@{}c@{}} \multirow{2}{*}{$\mathtt{TV \mhyphen on}$} & $\mathtt{\hspace{1mm} \myrightarrow{\mathtt{sound\mhyphen det.}} light \mhyphen on}$
          \end{tabular} & \multirow{2}{*}{$2$}  & TV's sound \textit{unintentionally} triggers two & \multirow{2}{*}{\xmark} & \multirow{2}{*}{\xmark} & \multirow{2}{*}{\xmark} \\
         & & \multirow{1}{*}{$\mathtt{\hspace{8.5mm} \myrightarrow{\mathtt{sound\mhyphen det.}} call \mhyphen user}$} &  & apps and causes light-on and call-user actions. & & & \\ \hline

         \multirow{3}{*}{$\mathtt{G_2}$}& \multirow{3}{*}{\textbf{$\mathtt{V_4}$}}& \multirow{1}{*}{$\mathtt{\hspace{44mm}\myrightarrow{\mathtt{sound\mhyphen det.}} light \mhyphen on}$}    & \multirow{3}{*}{$3$} & Aggregated sound from the AC, washer and dryer & \multirow{3}{*}{\xmark} & \multirow{3}{*}{\xmark} & \multirow{3}{*}{\xmark} \\
         && $\mathtt{Agg(AC \mhyphen on, washer \mhyphen on, dryer \mhyphen on)\myrightarrow{\mathtt{sound\mhyphen det.}} TV \mhyphen on}$ &  & \textit{unintentionally} triggers three apps and causes & & & \\ 
         && $\mathtt{\hspace{44mm}\myrightarrow{\mathtt{sound\mhyphen det.}} call \mhyphen user}$ &  &  light-on, TV-on and call-user actions. &&&\\ \hline

         \multirow{3}{*}{$\mathtt{G_3}$} & \multirow{3}{*}{$\mathtt{V_5}$}& \multirow{3}{*}{$\mathtt{Agg(bulb \mhyphen on, TV \mhyphen on)\myrightarrow{\mathtt{light\mhyphen det.}} light \mhyphen off}$}  & \multirow{3}{*}{$1$} & Aggregated light from the bulb and TV \textit{bypasses} & \multirow{3}{*}{\xmark} & \multirow{3}{*}{\xmark} & \multirow{3}{*}{\xmark}  \\ 
        & &    & & \multirow{1}{*}{bulb's intended influence and triggers an} & & & \\ 
        & &    & & \multirow{1}{*}{app that turns off the lights.} & & & \\ \hline

         \multirow{2}{*}{$\mathtt{DC_8}$} & \multirow{2}{*}{$\mathtt{V_6}$}& \multirow{2}{*}{$\mathtt{sleep \mhyphen mode \mhyphen activate\myrightarrow{\mathtt{sleep\mhyphen mode}} AC \mhyphen on \myrightarrow{\mathtt{temp < th \degree~F}} light \mhyphen on}$}  & \multirow{2}{*}{$1$} & Temperature decreases due to the AC's influence and & \multirow{2}{*}{\xmark} & \multirow{2}{*}{\cmark} & \multirow{2}{*}{\cmark}  \\ 
        & &  \multirow{1}{*}{}  & & turns on the bulb when the home mode is sleep. & & & \\ \hline
 
        \multirow{2}{*}{$\mathtt{DC_9}$} & \multirow{2}{*}{$\mathtt{V_7}$}& \multirow{1}{*}{$\mathtt{away \mhyphen mode \mhyphen activate\myrightarrow{\mathtt{away\mhyphen mode}} vacuum \mhyphen start \myrightarrow{\mathtt{motion \mhyphen det.}}}$}  & \multirow{2}{*}{$1$} & Temperature increases due to the heater's influence and & \multirow{2}{*}{\xmark} & \multirow{2}{*}{\cmark} & \multirow{2}{*}{\cmark}  \\ 
        & &  \multirow{1}{*}{$\mathtt{heater \mhyphen on \myrightarrow{\mathtt{temp > th \degree~F}} light \mhyphen on}$}  & & turns on the bulb when the home mode is away. & & & \\

        \thickhline
        \end{tabular}
        
    \end{threeparttable}}
\end{table*}

%% file: 0-CCS2022-FinalVersion/tables/placement_eval.tex
\begin{table}[t!]
    \centering
    \caption{The physical channels that cause policy violations after new device placement, and their tolerance to errors.} 
    \label{tab:placement_eval}
    \def\arraystretch{0.9}
    \setlength{\tabcolsep}{1.5em}
    \resizebox{\columnwidth}{!}{
    \begin{threeparttable}
        \begin{tabular}{|r|c|}
        \hline 
        \multicolumn{1}{|c|}{\textbf{Physical Channel}} & \textbf{Tolerance to Distance Error} \\ \hline 
        $\mathtt{washer \mhyphen on}$    $\mathbin{\stackon[-0.3pt]{\squigs{2}\rsquigend}{\scriptscriptstyle\text{\,}}}$ $\mathtt{sound \mhyphen detected}$   & $20\%$\\ \hline 
        $\mathtt{dryer \mhyphen on}$    $\mathbin{\stackon[-0.3pt]{\squigs{2}\rsquigend}{\scriptscriptstyle\text{\,}}}$ $\mathtt{sound \mhyphen detected}$ & $>50\%$ \\ \hline 
        $\mathtt{TV \mhyphen on}$    $\mathbin{\stackon[-0.3pt]{\squigs{2}\rsquigend}{\scriptscriptstyle\text{\,}}}$ $\mathtt{sound \mhyphen detected}$ & $>50\%$ \\ \hline  
        \end{tabular}
    \end{threeparttable}}
\end{table}

%% file: 0-CCS2022-FinalVersion/text/discussion.tex
\section{Discussion \& Limitations}
\label{sec:discussion}

\shortsectionBf{Mitigating the Policy Violations.}
\system identifies policy violations and presents users with a report that details the violation's root cause. However, there is a need to mitigate the policy violations to ensure the safe and secure operation of the smart home. We discuss three methods for mitigation, ($1$) patching the app code, ($2$) device placement changes, and ($3$) removal of apps.

The first mitigation technique is patching the app code to block its commands if the app is triggered due to an unintended influence.
This technique adds a code block that guards an app's action with a predicate conditioned on the devices that unintentionally influence the app's sensor event.
If a device is unintentionally influencing a channel, the predicate becomes false, preventing the app from issuing its command. 
For instance, the {\small{$\mathtt{V_4}$}} violation is prevented by adding a predicate to the apps conditioned on {\small{$\mathtt{sound \mhyphen detected}$}}. The predicate blocks apps' actions if the AC, washer, and dryer are simultaneously turned on, preventing the unintended interactions.

Second, we recommend users increase the distance between the actuator and sensor to prevent policy violations.  
This is because a command's influence on sensor readings monotonically decreases as the distance between the actuator and sensor increases~\cite{voudoukis2017inverse,zhivov2001principles}. 
For instance, operating the robot vacuum away from the motion sensor (\eg by setting `keep out zones') prevents the motion from {\small{$\mathtt{vacuum \mhyphen on}$}} from triggering events.
Lastly, removing one or more apps involved in the physical interaction prevents a violation.
A user may prefer this method if an app is not critically needed and other mitigation techniques are not feasible.

\input{0-CCS2022-FinalVersion/tables/mitigation}
For the $16$ identified policy violations, we applied the above mitigation methods and evaluated their effectiveness. 
Table~\ref{tab:mitigation} shows the number of policy violations prevented by our mitigation methods. 
First, patching the app code prevents $14$ out of $16$ violations as it inserts predicates that guard the unintended physical interactions.
For instance, {\small{$\mathtt{V_{1}}$}} is patched by adding a condition to the apps triggered by the motion detected event. The condition checks whether the robot vacuum is not on before sending the app's commands.
Second, changing the device placement prevents all violations. We validate this through in-home experiments with increased distance between the devices in the policy violations. 
For instance, increasing the distance from the garbage disposal to the sound sensor prevents {\small{$\mathtt{V_{2}}$}} as {\small{$\mathtt{garbage\mhyphen disposal \mhyphen on}$}}'s sound cannot reach the sensor.
Finally, removing the apps involved in the physical interactions prevents all violations, \eg removing a single app from the three apps that cause the {\small{$\mathtt{V_{6}}$}} violation prevents it.

We note that our mitigation methods may prevent desired actions while eliminating dangerous app interactions.
First, code patching blocks actions while actuators that unintentionally influence a physical channel are turned on, yet, a user may desire to issue those actions.
In such cases, the user can manually activate them and respond to the app interactions. 
Second, changing a device's placement may be inconvenient for the user, and it may cause other policy violations.
However, \system can identify new policy violations by updating its distance parameters and running its security analysis module.
Lastly, removing an app eliminates all of its interactions; however, users may not desire to remove the apps they need.
In future work, we will conduct user studies to learn how users perceive the mitigation methods and investigate advanced techniques for automated patching.
For instance, we will explore automated distance range discovery through parameter mining~\cite{hoxha2018mining} to find the specific distance ranges between actuators and sensors that can prevent all policy violations. 

\shortsectionBf{Manual Effort Required.}
\system requires users' effort in determining the distance parameter in the \comp and the {\small{$\mathtt{Int/UnInt}$}} labels for intent-based policies.
First, the users need to confirm that the distances found from RSSI-based localization are correct and provide the distances manually if necessary.
This effort is not a significant burden for users since they can provide approximate distances. This is because \system can tolerate small errors in the distance parameter, as shown in our evaluation.
Second, although \system generates {\small{$\mathtt{Int/UnInt}$}} labels between commands and apps, the users may have different intentions than the generated ones.
Therefore, they need to check the labels and update them if necessary based on their intended use of the actuators and apps. 

\shortsectionBf{Environmental Noise.}
The environment that the devices operate in may influence the physical channels and impact the app interactions.
As \system leverages SI to tune \comp parameters, it integrates various environmental impacts such as room layout and furniture. 
Yet, human activities and uncontrolled environmental noise may also influence sensor measurements.
To measure the impact of human activities, we conducted additional experiments while a user was cooking and exercising. We have found that the users do not introduce detectable changes to sensors.
We have shown in in-home validation experiments that uncontrolled noise causes the violations to occur at slightly different times.

%% file: 0-CCS2022-FinalVersion/tables/mitigation.tex
\begin{table}[t!]
    \centering
    \caption{Mitigated policy violations with different methods.} 
    \label{tab:mitigation}
    \setlength{\tabcolsep}{1.4em}
    \resizebox{\columnwidth}{!}{
    \begin{threeparttable}
        \begin{tabular}{l|c|c|c|c|}
        \cline{2-5}
            & \multicolumn{4}{c|}{\textbf{Policy}}                   \\ \hline
            \multicolumn{1}{|l|}{\textbf{Mitigation Method}} & \textbf{G1} & \textbf{G2} & \textbf{G3} & \textbf{DC} \\ \hline
            \multicolumn{1}{|l|}{Patching the App Code}      & $10/10$         & $3/3$         & $1/1$         & $0/2$         \\ \hline
            \multicolumn{1}{|l|}{Device Placement Changes}   & $10/10$         & $3/3$         & $1/1$         & $2/2$         \\ \hline
            \multicolumn{1}{|l|}{Removal of Apps}            & $10/10$         & $3/3$         & $1/1$         & $2/2$         \\ \hline
        \end{tabular}
    \end{threeparttable}}
\end{table}

%% file: 0-CCS2022-FinalVersion/text/related.tex
\section{Related Work}
\label{sec:related}
\input{0-CCS2022-FinalVersion/tables/comparison}

In Table~\ref{tab:comp}, we compare \system with several recent approaches that focus on identifying the vulnerabilities that IoT app interactions present.
These approaches can be classified into two categories: run-time policy enforcement and static analysis for IoT apps.

\shortsectionBf{Run-time Policy Enforcement.}
IoTGuard~\cite{celik2019iotguard} instruments apps to build dynamic models and enforces policies at run-time.
Yet, it cannot correctly identify physical interactions since its models do not include the commands' physical influences.

IoTSafe~\cite{ding2021iotsafe} collects actuator and sensor traces to identify physical interactions and builds physical models for continuous physical channels to predict incoming policy violations at run-time. 
However, it cannot identify intent-based violations as its policies are only defined based on the use cases of devices, and it cannot determine the specific command that influences a physical channel from examining sensor measurements at run-time.
Additionally, IoTSafe does not consider distance in its models and flags incorrect violations or fails to detect a violation when a device's placement is changed.
These systems motivate the need for \system, which can precisely identify dangerous physical interactions before the smart home's run-time operation.

\shortsectionBf{Static Analysis of IoT Apps.}
Existing static analysis systems do not model apps' complex physical behavior. Instead, they build individual physical channel mappings, generate naive device behavioral models~\cite{bu2018systematically,wang2019charting}, or use natural language processing~\cite{ding2018safety} to infer interacting apps.
Thus, they identify limited physical interactions and lead to false positives. 
As presented in Table~\ref{tab:comp}, \system is the first to integrate the complex physical properties of commands and sensor events into the source code of IoT apps (``Physical Channel Properties'' Columns). 
Additionally, their validation techniques cannot readily be used to verify physical interactions as apps exhibit both discrete and continuous behaviors. 
In contrast, \system extends optimization-guided falsification for scalable policy validation.

%% file: 0-CCS2022-FinalVersion/tables/comparison.tex
\begin{table}[t!]
\caption{Comparison of \system with IoT security systems.}
\label{tab:comp}
{\footnotesize{
    \centering
    \setlength{\tabcolsep}{0.3em}
    \resizebox{\columnwidth}{!}{
    \begin{threeparttable}
        \begin{tabular}{c|c|c|c|c|c|c|c|} 
         \cline{2-8}
         & \multicolumn{5}{c|}{\textbf{Physical Channel Properties}} & \multicolumn{2}{c|}{\textbf{Security Analysis}}   \\  \hline
        \multicolumn{1}{|c|}{\multirow{2}{*}{\textbf{System}}} & \multirow{2}{*}{\textbf{Dist.}} & \multirow{2}{*}{\textbf{Agg.}} & \multirow{2}{*}{\textbf{Dep.}} & \textbf{Labels}  & \multirow{2}{*}{\textbf{Composition}}& \textbf{Time} & \textbf{Policy} \\ 
        \multicolumn{1}{|c|}{} & & &  & \textbf{(Int/UnInt)}  & \textbf{} & \textbf{Constraints} & \textbf{Validation$^\dagger$} \\ \hline 
         \multicolumn{1}{|c|}{IotGuard$^*$~\cite{celik2019iotguard}} & \xmark & \xmark &  \xmark & \xmark  & \xmark & \xmark & \texttt{RA}  \\ \hline
         
         \multicolumn{1}{|c|}{IoTSafe$^*$~\cite{ding2021iotsafe}} & \xmark & \cmark &  \cmark & \xmark  & \xmark & \xmark  & \texttt{RP} \\ \hline
         
        \multicolumn{1}{|c|}{MenShen~\cite{bu2018systematically}} & \xmark & \xmark & \xmark  &  \xmark  & \xmark & \xmark  &   \texttt{RA} \\ \hline
         \multicolumn{1}{|c|}{iRuler~\cite{wang2019charting}} & \xmark  & \xmark & \xmark  &  \xmark   & \xmark  & \xmark & \texttt{RL}  \\ \hline
         \multicolumn{1}{|c|}{IoTCom~\cite{alhanahnah2020scalable}} & \xmark & \xmark &  \xmark & \xmark  & \xmark & \xmark  & \texttt{MC}  \\ \hline
         \multicolumn{1}{|c|}{IoTMon~\cite{ding2018safety}} & \xmark & \xmark &  \xmark & \xmark  & \xmark & \xmark & \texttt{N/A}  \\ \hline

         \multicolumn{1}{|c|}{\system} & \cmark & \cmark  & \cmark & \cmark  & \cmark & \cmark & \texttt{F}  \\ \hline
        \end{tabular}
        $*$ IoTGuard and IoTSafe are run-time policy enforcement systems. 
        
        $\dagger$ \texttt{RA}: Reachability Analysis, \texttt{RL}: Rewriting Logic, \texttt{MC}: Model Checking, \texttt{F}: Optimization-guided Falsification, \texttt{RP}: Run-time Prediction. 
    \end{threeparttable}
    }}}
\end{table}

%% file: 0-CCS2022-FinalVersion/text/conclusion.tex
\section{Conclusions}
\label{sec:conclusion}

We introduce \system, which identifies the physical channel vulnerabilities in smart homes.
\system combines static app analysis with system identification to precisely model the composite physical behavior of apps and uses falsification to validate identified physical channel policies.
Our evaluation in a real house demonstrates that many apps interact over physical channels, and \system efficiently and effectively identifies all policy violations.
This paper is an important step forward in achieving the compositional safety and security of an IoT system's physical behavior.

%% file: 0-CCS2022-FinalVersion/text/appendix.tex
\appendix

\section{Appendix Guide}
\label{sec:app_guide}
This Appendix provides the information necessary to reproduce our results.
In Appendix~\ref{appendix:deviceRules}, we present the identified device-centric policies. 
In Appendix~\ref{appendix:fidelity}, we describe the \comp fidelity experiments and present our numerical results.
In Appendix~\ref{appendix:simulink}, we present an example \comp developed with Simulink.
In Appendix~\ref{sec:apps_experiments}, we detail the apps considered in our evaluation. 
In Appendix~\ref{sec:indModelFormulae}, we present the flow functions of studied actuation commands and sensor events.
In Appendix~\ref{appendix:details}, we present the influence of actuation commands on physical channels, the distances between the actuators and sensors, and the sensor sensitivity levels.
We note that \system contains all tools necessary to extend the \comp to other apps and allows for replacing the flow functions of studied actuation commands and sensor events with the ones that may be a better contextual fit.

\section{Device-centric Policies}
\label{appendix:deviceRules}

We present the identified device-centric policies in Table~\ref{tab:device-policy-full}.

\input{0-CCS2022-FinalVersion/tables/device-centric-full}

\section{CPEM Fidelity Experiments}
\label{appendix:fidelity}

We evaluate the fidelity of \comp compared to the actual devices. 
We select all actuation command and sensor pairs where the sensor measures the command's influence. 
We set the distance in each pair to $\mathtt{0.5}$ m - $\mathtt{2.5}$ m with $\mathtt{0.5}$ m intervals.
We collect \apem and real device traces for each distance to evaluate their fidelity.
We compute the $(\tau,\epsilon)$-closeness of actual device traces with the \apem traces to measure \comp fidelity.
%
Let $\mathtt{x}$ be a \apem's traces, and $\mathtt{y}$ be the real device traces generated with the same inputs. 
Given $\mathtt{T} \in \mathbb{R}_+$, and $(\tau,\epsilon) \geq 0$, we determine $\mathtt{x}$ and $\mathtt{y}$ are $(\tau,\epsilon)$-close if for all $\mathtt{t \in x}$, $\mathtt{t \leq T}$, there exists $\mathtt{s \in y}$ where $\mathtt{|t-s| \leq \tau}$, and $\mathtt{|x(t)-y(s)| \leq \epsilon}$, and  for all $\mathtt{t \in y}$, $\mathtt{t \leq T}$, there exists $\mathtt{s \in x}$ where $\mathtt{|t-s| \leq \tau}$ and $\mathtt{|y(t)-x(s)| \leq \epsilon}$.

\input{0-CCS2022-FinalVersion/tables/conformance_experiment}

Table~\ref{tab:conformance_exp} details the $(\tau,\epsilon)$-closeness of actual device traces with the \apem traces. 
Each row includes the mean and standard deviation of the timing difference $(\tau)$ and deviation score $(\epsilon)$ over different distances.
We observe slight deviations in temperature, illuminance, sound, and humidity sensor \apem outputs and no deviation in motion sensor \apem outputs.
This shows that our actuator and sensor \apems have high fidelity with actual device traces. 
The slight deviations are due to the uncertainties in the environmental factors (\eg the sunlight amount in the room~\cite{elishakoff2000whys}) that impact actual device traces.

\begin{figure}[t]
\begin{subfigure}{.23\textwidth}
  \centering
  \includegraphics[width=\linewidth]{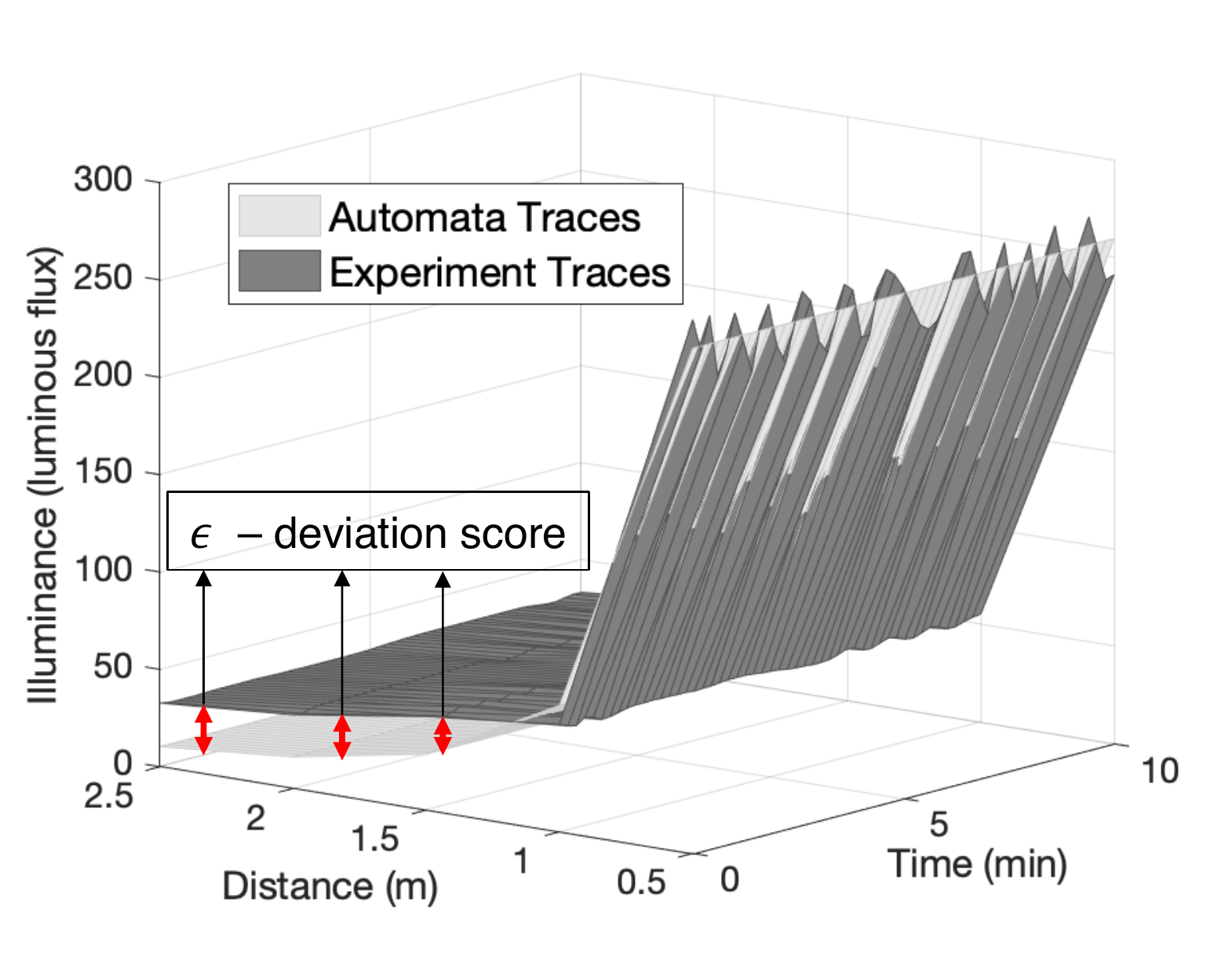}  
  \caption{}
  \label{fig:lamp_conformance}
\end{subfigure}
\begin{subfigure}{.23\textwidth}
  \centering
  \includegraphics[width=\linewidth]{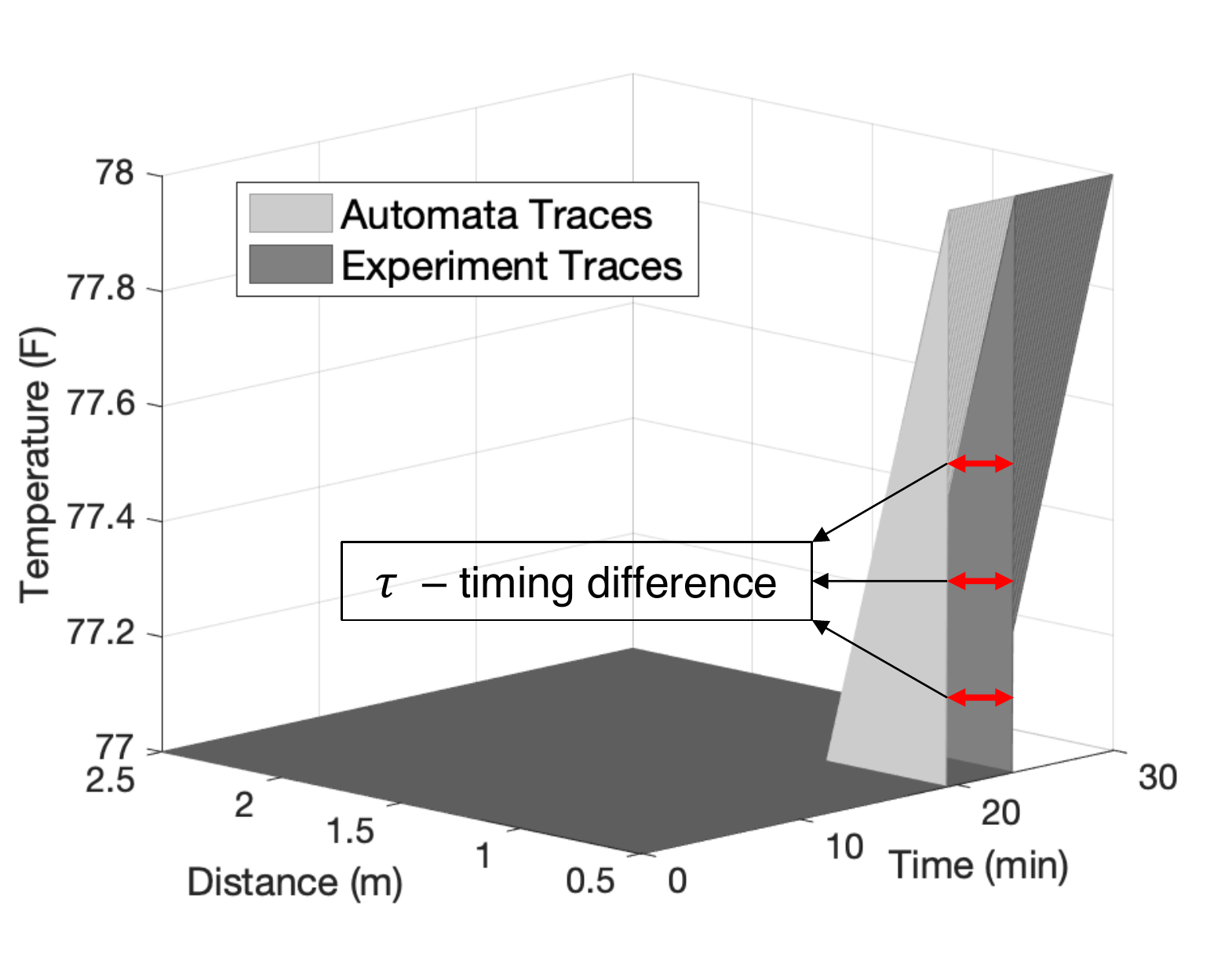}  
  \caption{}
  \label{fig:oven_conformance}
\end{subfigure}
\caption{{\apem and device traces: (a) {\small{$\mathtt{bulb \mhyphen on}$}} changes the illuminance sensor readings and (b) {\small{$\mathtt{oven \mhyphen on}$}} changes the temperature sensor readings.}}
\label{fig:conformance}
\end{figure}

To illustrate, Figure~\ref{fig:conformance} plots the traces from \apems and actual device experiments of {\small{$\mathtt{light \mhyphen bulb \mhyphen on}$}}'s influence on the illuminance sensor and {\small{$\mathtt{oven \mhyphen on}$}}'s influence on the temperature sensor.
We found the {\small{$\mathtt{bulb \mhyphen on}$}} \apem deviates on average $\mathtt{\approx 20~lux}$ from device traces. 
Additionally, {\small{$\mathtt{oven \mhyphen on}$}} increases the temperature sensor readings by $\mathtt{1.8 \degree F}$ in both \apem and actual device traces at $\mathtt{0.5}$ meters. 
The \apem yields an increase at minute $\mathtt{19.6}$, but the increase in the device traces occurs at minute $\mathtt{23.8}$. 
This leads to $\mathtt{\tau = 0.8\pm1.9}$ timing difference and $\mathtt{\epsilon = 0}$ deviation score. 
As detailed in Sec.~\ref{subsec:effectiveness}, these lead to safe over-approximations in detecting violations at slightly different times.

\section{Simulink CPEM Example}
\label{appendix:simulink}
\begin{figure}[t!]
    \centering
    \includegraphics[width=.9\linewidth]{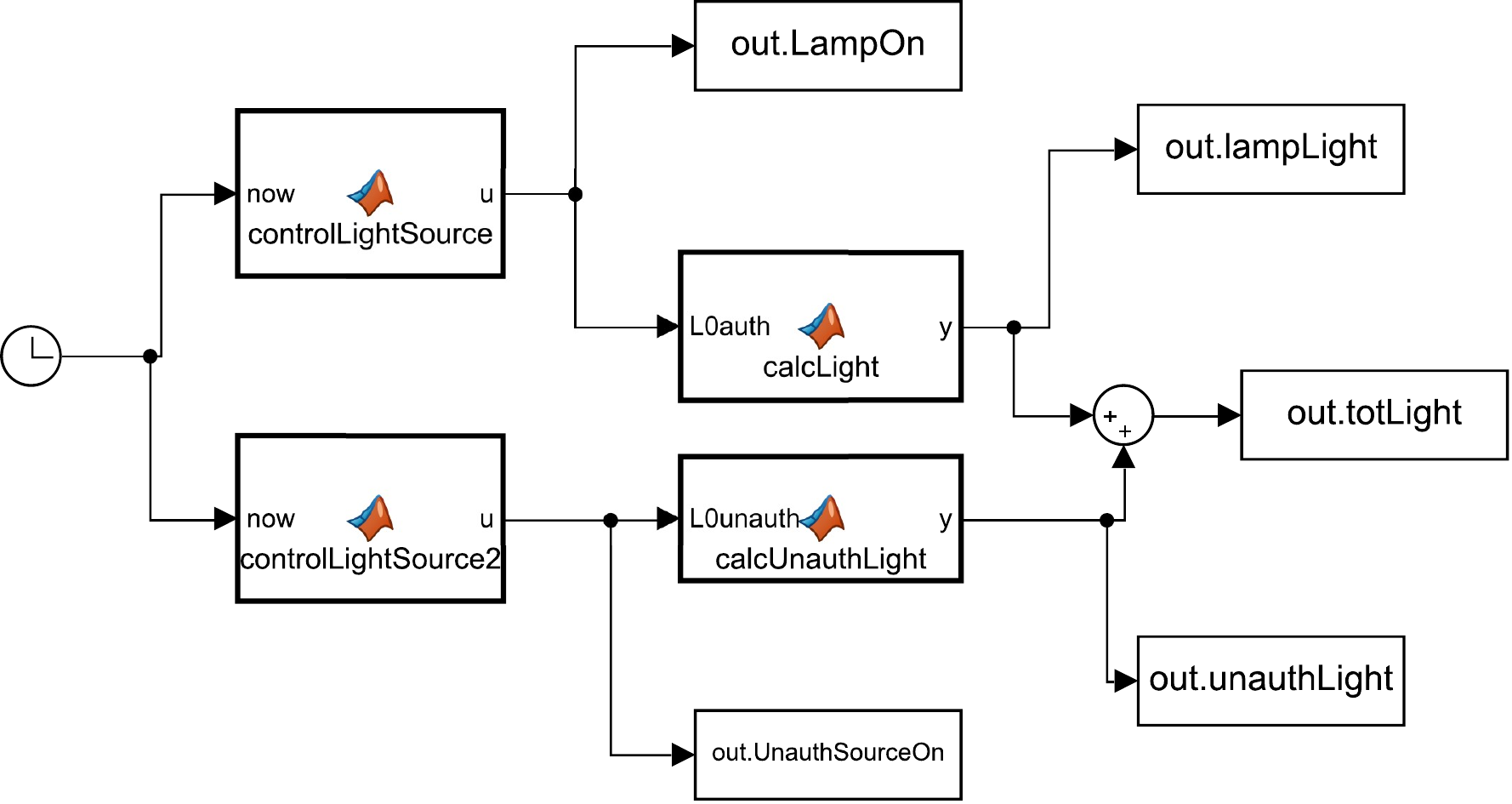}
    \caption{Simulink \comp of the {\small{$\mathtt{light \mhyphen bulb \mhyphen on}$}}, {\small{$\mathtt{TV \mhyphen on}$}} actuation commands, and the {\small{$\mathtt{light \mhyphen detected}$}} event.}
    \label{fig:simulink}
\end{figure}

The \comp of {\small{$\mathtt{light \mhyphen bulb \mhyphen on}$}}, {\small{$\mathtt{TV \mhyphen on}$}} commands, and the {\small{$\mathtt{light \mhyphen detected}$}} event is presented in Figure~\ref{fig:simulink}. The actuation command and sensor event \apems (including their flow functions) are implemented with Matlab functions in the Simulink boxes. 
Our implementation can be easily migrated to other software such as GNU Octave~\cite{GNUOctaveWebsite}, Scilab~\cite{ScilabWebsite}, and SpaceEx~\cite{SpaceExWebsite} with third-party code converters.

\section{Apps Deployed in the Experiments}
\label{sec:apps_experiments}

The full list of the apps considered in our experiments is presented in Table~\ref{tab:apps}.

\input{0-CCS2022-FinalVersion/tables/apps_experiment}

\section{PEM Flow Functions}
\label{sec:indModelFormulae}
To determine the flow functions of \apems, we first analyze the physical channels each sensor event observes~\cite{jackson1996sensor,voudoukis2017inverse}. 
We then study how actuation commands influence these channels and how they diffuse in a way that changes the sensor readings~\cite{bloomfield2007everything,bukowski2003performance,hancock20061,lawrence2005relationship,salazar2003thermal,zhivov2001principles}. 
Here, we present the equations that rely on the laws of physics, which are integrated into our \apems as flow functions. 
Different equations that can capture more detailed influences of actuation commands on physical channels can be easily integrated by updating the corresponding flow functions. 

\shortsectionBf{Temperature.} Appliances' influences on temperature are due to hot surfaces diffusing heat to their surroundings.
Therefore, the heat diffusion equation from a point source~\cite{hancock20061} (which is based on a partial differential equation) is used as their flow function. 

\begin{equation}
 \frac{\partial \mathtt{T}}{\partial \mathtt{t}} = \alpha \frac{\partial^2\mathtt{T}}{\partial \mathtt{x}^2} 
\end{equation}

\begin{equation}
\mathtt{T}(\mathtt{e},0)=\mathtt{T_0}
\end{equation}
\begin{equation}
\mathtt{T}(0,0)=\mathtt{T_s}
\end{equation}

In the formula, $\mathtt{T}$ is the temperature in $\degree \mathtt{K}$ , $\mathtt{x}$ is the distance from the actuator in meters, $\alpha$ is the thermal diffusivity constant (in $\mathtt{m}^2/\mathtt{s}$), $\mathtt{e}$ is the maximum distance from the source, and $\mathtt{T_s}$ is the temperature of the source. We set the thermal diffusivity to $2.2 \cdot 10^{-5}$ $\mathtt{m}^2/\mathtt{s}$ as a constant~\cite{hilsenrath1955tables,salazar2003thermal}.

On the other hand, HVAC systems (\eg heater and AC) influence temperature with an airflow to ensure quick heat dissipation (due to convection).
In particular, the airflow of the HVAC systems enables them to uniformly influence different locations~\cite{glicksman1997thermal}.
Therefore, we construct their flow functions through ordinary differential equations~\cite{goebel2009hybrid}.

\begin{equation}
 \frac{d \mathtt{T}}{d \mathtt{t}} =  \mathtt{T}_\Delta \cdot \mathtt{q}
\end{equation}
\begin{equation}
\mathtt{T}(0)=  \mathtt{T_0} 
\end{equation}

In the equation, $\mathtt{T}_0$ is the initial temperature, $\mathtt{T}_\Delta$ is the impact from the HVAC system based on its power (positive for the heater, negative for the AC), and $\mathtt{q}$ denotes the system's actuation command ($\mathtt{on}$ or $\mathtt{off}$). Although these devices cause an airflow in the room in order to enable uniform temperature, the appliances still generate heat through diffusion and influence the sensor measurements. Therefore, both diffusion and convection play a critical role in  temperature sensor measurements.

\shortsectionBf{Relative Humidity.} 
The relative humidity is defined as the ratio of the existing water vapor in the air to the maximum capacity of water vapor that can exist in the air~\cite{lawrence2005relationship}. 

\begin{equation}
\mathtt{RH}\% = \frac{\mathtt{w}}{\mathtt{w_s}}\times 100
\end{equation}

In the formula, $\mathtt{w}$ represents the water vapor in the air, and $\mathtt{w_s}$ represents the maximum capacity of water vapor the air can contain. $\mathtt{w_s}$ exponentially depends on the environmental temperature, related to the Clausius–Clapeyron equation. Therefore, we leverage existing experiments~\cite{van2006moisture} to fit an exponential function to compute $\mathtt{w_s}$ based on temperature. In particular, this function is defined as:

\begin{equation}
\mathtt{w_s} = 2.6055\times e^{0.0262\times \mathtt{T}}
\end{equation}

\noindent where $\mathtt{T}$ represents the environmental temperature in $\mathtt{\degree F}$. 

We implement the actuation command \apem's flow functions as ordinary differential equations that generate or reduce water content in the air.

\shortsectionBf{Smoke.} We consider an ionization-based smoke detector that detects the presence of smoke by filtering air through an ionization chamber. When smoke particles enter the chamber, conductivity decreases, and smoke presence is detected~\cite{bukowski2003performance}. The threshold of these sensors is in obscuration density per meter ($\mathtt{OD/m}$), where a typical sensor's sensitivity is $0.02$ $\mathtt{OD/m}$, corresponding to $13$  $\mathtt{mg/m^3}$ smoke density~\cite{bukowski2003performance,lee1977physical}.
Gas particles (such as the particles in smoke) move very fast through the air at room temperature~\cite{bloomfield2007everything}. 
Therefore, we define the flow function of actuation command \apems as an ordinary differential equation.

\shortsectionBf{Illuminance.} The influence of light is instant. Therefore, it is modeled with an algebraic equation. Its influence follows the inverse square law, as shown below~\cite{voudoukis2017inverse}. 

\begin{equation}
\mathtt{I_x} = \frac{\mathtt{I_s}}{4\times\pi \times \mathtt{x}^2}
\end{equation}

In the formula, $\mathtt{I_s}$ denotes the luminosity flux of the source.

\shortsectionBf{Sound.} 
Sound is modeled with an algebraic equation due to its high diffusion speed. Sound intensity is modeled with the inverse square law~\cite{voudoukis2017inverse}. Therefore, the sound pressure, which is the quantity the sensors measure, is represented with the following formula.

\begin{equation}
\mathtt{SP_2} = \mathtt{SP_1 + 20}\times \mathtt{log}_{10}(\frac{\mathtt{x_1}}{\mathtt{x_2}})
\end{equation}

Sound pressure $\mathtt{SP_2}$ at distance $\mathtt{x_2}$ can be calculated in decibels ($\mathtt{dB}$) with this formula, where $\mathtt{SP_1}$ is the sound pressure level at distance $\mathtt{x_1}$ (the standard value of $\mathtt{x_1}$ is $\mathtt{1}$ meter)~\cite{winer2012audio}. When the distance is doubled, $\mathtt{SP_2}$ decreases by $6$ $\mathtt{dB}$.

\shortsectionBf{Motion.} There are different types of motion sensors available on the market. For instance, accelerometers detect motion based on the acceleration generated by the source, PIR sensors detect motion based on the infrared heat map of an environment, and laser-based sensors detect motion by generating an invisible laser between two devices and detecting the cut-offs. We consider PIR motion sensors and model them through their range to detect motion. If the actuators' movement rate exceeds a threshold and the distance is close enough, motion is detected.

\begin{figure}[t!]
    \centering
    \includegraphics[width=\linewidth]{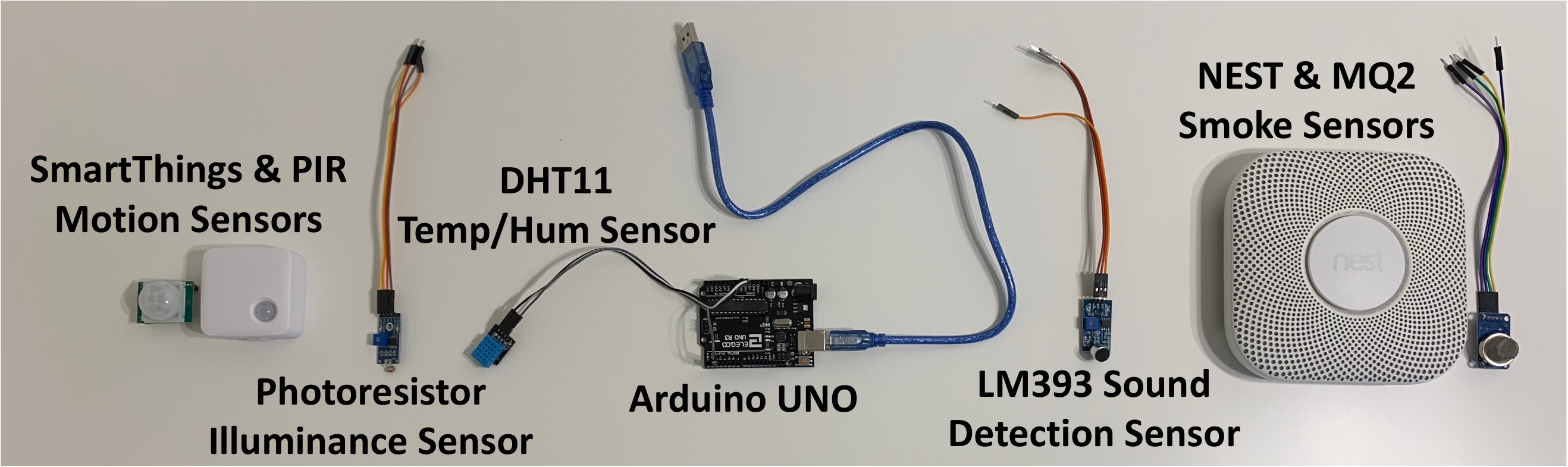}
    \caption{Sensors installed in the house (See Figure~\ref{fig:house}).} 
    \label{fig:sensors}
\end{figure}

\section{Device Specifications}
\label{appendix:details}

Table~\ref{tab:actuator_detail} presents the actuator configurations, influences of the actuators on sensors, and distances to the sensors they influence.

For sensors (presented in Figure~\ref{fig:sensors}), sensitivity values, and thresholds for detection (\eg for sound, illuminance, smoke) are set as follows: (i) temperature sensor - $\pm 1.8 \degree\mathtt{F}$, (ii) humidity sensor - $\pm \mathtt{2\%}$, (iii) smoke sensor - $0.02~\mathtt{OD/m} \approx 13~\mathtt{mg/m^3}$, (iv) illum. sensor - $\mathtt{50}$ lux, (v) sound sensor - $55$ dB, and (vi) motion sensor - $\mathtt{1}$ m.

\input{0-CCS2022-FinalVersion/tables/actuator_details}

%% file: 0-CCS2022-FinalVersion/tables/device-centric-full.tex
\begin{table}[t!]
\caption{Device-centric security policies.}
    \label{tab:device-policy-full}
    \centering
    \def\arraystretch{1.05}
    \resizebox{\columnwidth}{!}{
    \begin{threeparttable}
        \begin{tabular}{|c|c|c|} 
         \thickhline
         \textbf{ID} & \textbf{Policy Description} & \textbf{Formal Representation}  \\ \thickhline
         \multirow{2}{*}{$\mathtt{DC_1}$} & The sprinkler system must go off   &  \multirow{2}{*}{$\mathtt{\square(fire\rightarrow\Diamond_{[0,{t}]}sprinkler \mhyphen on)}$}  \\ 
         & within $\mathtt{t}$ seconds when there is a fire. & \\ \hline
         \multirow{2}{*}{$\mathtt{DC_2}$} & When the home is in the away mode, & \multirow{2}{*}{$\mathtt{\square(mode \mhyphen away \rightarrow window \mhyphen close)}$}  \\ 
         & the window must be closed. & \\ \hline
         \multirow{2}{*}{$\mathtt{DC_3}$} & A device must not open, then close and  &  \multirow{2}{*}{$\mathtt{\square\neg(on\wedge\Circle\Diamond_{[0,{t}]}( off\wedge \Circle\Diamond_{[0,{t}]} on))}$} \\
         & then reopen (actuation loop) within $\mathtt{t}$ seconds. &   \\  \hline

         \multirow{2}{*}{$\mathtt{DC_4}$} & The window must not open  &  \multirow{2}{*}{$\mathtt{\square (heater \mhyphen on\rightarrow \neg window \mhyphen open) }$}  \\ 
         & when the heater is on. & \\ \hline

         \multirow{2}{*}{$\mathtt{DC_5}$} & The alarm must go off within $\mathtt{t}$ seconds  & \multirow{2}{*}{$\mathtt{\square(smoke \mhyphen detected \rightarrow\Diamond_{[0,{t}]}alarm \mhyphen on)}$}  \\
         &after smoke is detected. & \\\hline

        \multirow{2}{*}{$\mathtt{DC_6}$} & The main door must not be left  &   \multirow{2}{*}{$\mathtt{\Diamond_{[0,{t}]}door \mhyphen lock}$} \\
         & unlocked for more than $\mathtt{t}$ seconds. & \\ \hline
         
         \multirow{2}{*}{$\mathtt{DC_7}$} & The window must not be open &\multirow{2}{*}{$\mathtt{\square (AC \mhyphen on\rightarrow \neg window \mhyphen open) }$}   \\ 
         & when the AC is on. & \\ \hline 
         
         \multirow{2}{*}{$\mathtt{DC_8}$} & The lights must be turned off when  & \multirow{2}{*}{$\mathtt{\square(mode \mhyphen sleep \rightarrow light \mhyphen off)}$}  \\ 
         & the home is in the sleep mode. & \\ \hline 
         
         \multirow{2}{*}{$\mathtt{DC_{9}}$} & The door must always be locked and lights  & \multirow{2}{*}{$\mathtt{\square(mode\mhyphen away \rightarrow door \mhyphen lock \wedge light \mhyphen off)}$}  \\ 
         & must be off when the home is in the away mode. & \\ \hline 
         
         \multirow{2}{*}{$\mathtt{DC_{10}}$} & The TV must be turned off when & \multirow{2}{*} {$\mathtt{\square((mode \mhyphen away \vee mode \mhyphen sleep ) \rightarrow TV \mhyphen off)}$} \\ 
         & the home is in the away or sleep mode. & \\ \thickhline

        \end{tabular}
              \end{threeparttable}}
\end{table}

%% file: 0-CCS2022-FinalVersion/tables/conformance_experiment.tex
\begin{table}[t!]
    \centering
     \caption{$\mathtt{(\tau, \epsilon)}$-closeness of \comp and actual IoT devices.}
    \label{tab:conformance_exp}
    \setlength{\tabcolsep}{0.2em}
    \def\arraystretch{0.9}
    \resizebox{\columnwidth}{!}{
    \begin{threeparttable}
        \begin{tabular}{c|c|c|c|c|c|c|c|c|c|c|c|c|} 
         \cline{2-13}
         &  \multicolumn{2}{c|}{\textbf{Temperature}} & \multicolumn{2}{c|}{\textbf{Illum.}} & \multicolumn{2}{c|}{\textbf{Sound}} & \multicolumn{2}{c|}{\textbf{Humidity}} & \multicolumn{2}{c|}{\textbf{Motion}} & \multicolumn{2}{c|}{\textbf{Smoke}}   \\ \hline
         \multicolumn{1}{|c|}{\textbf{Actuator}} & \textbf{$\mathtt{\tau}$ } & \textbf{$\mathtt{\epsilon}$} & \textbf{$\mathtt{\tau}$} & \textbf{$\mathtt{\epsilon}$} & \textbf{$\mathtt{\tau}$} & \textbf{$\mathtt{\epsilon}$} & \textbf{$\mathtt{\tau}$} & \textbf{$\mathtt{\epsilon}$} & \textbf{$\mathtt{\tau}$} & \textbf{$\mathtt{\epsilon}$} & \textbf{$\mathtt{\tau}$} & \textbf{$\mathtt{\epsilon}$} \\ \hline \hline

          \multicolumn{1}{|c|}{Heater}& $\mathtt{8.2 \pm 4.3}$ & $\mathtt{0}$ & \multicolumn{2}{c|}{$\mathtt{N/A}$} & \multicolumn{2}{c|}{$\mathtt{N/A}$} & $\mathtt{5.2 \pm 1.8}$ & $\mathtt{3.2 \pm 1.1}$ & \multicolumn{2}{c|}{$\mathtt{N/A}$} & \multicolumn{2}{c|}{$\mathtt{N/A}$} \\ \hline

          \multicolumn{1}{|c|}{Oven} & $\mathtt{0.8 \pm 1.9}$ & $\mathtt{0}$ & \multicolumn{2}{c|}{$\mathtt{N/A}$} & \multicolumn{2}{c|}{$\mathtt{N/A}$} & $\mathtt{1.3 \pm 2.9}$ & $\mathtt{0}$ & \multicolumn{2}{c|}{$\mathtt{N/A}$} & \multicolumn{2}{c|}{$\mathtt{N/A}$}  \\ \hline

          \multicolumn{1}{|c|}{Pressure Cooker} & $\mathtt{1 \pm 2.1}$ & $\mathtt{0.4 \pm 0.8}$ & \multicolumn{2}{c|}{$\mathtt{N/A}$} & \multicolumn{2}{c|}{$\mathtt{N/A}$} & $\mathtt{1 \pm 2.2}$ & $\mathtt{0.4 \pm 0.9}$ &\multicolumn{2}{c|}{$\mathtt{N/A}$} & \multicolumn{2}{c|}{$\mathtt{N/A}$} \\ \hline

          \multicolumn{1}{|c|}{Coffee maker} & $\mathtt{0}$ & $\mathtt{0}$ & \multicolumn{2}{c|}{$\mathtt{N/A}$} & \multicolumn{2}{c|}{$\mathtt{N/A}$} & $\mathtt{0}$ & $\mathtt{0}$ &\multicolumn{2}{c|}{$\mathtt{N/A}$} & \multicolumn{2}{c|}{$\mathtt{N/A}$} \\ \hline

          \multicolumn{1}{|c|}{DeHumidifier} & \multicolumn{2}{c|}{$\mathtt{N/A}$}  & \multicolumn{2}{c|}{$\mathtt{N/A}$} & \multicolumn{2}{c|}{$\mathtt{N/A}$} & $\mathtt{0}$ & $\mathtt{0}$ &\multicolumn{2}{c|}{$\mathtt{N/A}$} & \multicolumn{2}{c|}{$\mathtt{N/A}$}\\ \hline
         
          \multicolumn{1}{|c|}{Clothes Washer} & \multicolumn{2}{c|}{$\mathtt{N/A}$} & \multicolumn{2}{c|}{$\mathtt{N/A}$} & $\mathtt{0}$ & $\mathtt{0.2 \pm 0.4}$ & $\mathtt{0}$ & $\mathtt{0}$ & \multicolumn{2}{c|}{$\mathtt{N/A}$} &\multicolumn{2}{c|}{$\mathtt{N/A}$} \\ \hline
          
          \multicolumn{1}{|c|}{Dryer} & $\mathtt{0}$ & $\mathtt{1.1 \pm 2.4}$ & \multicolumn{2}{c|}{$\mathtt{N/A}$} &  $\mathtt{0}$ & $\mathtt{0.2 \pm 0.4}$ & $\mathtt{0}$ & $\mathtt{1.2 \pm 2.7}$ & \multicolumn{2}{c|}{$\mathtt{N/A}$} & \multicolumn{2}{c|}{$\mathtt{N/A}$} \\ \hline
          
         \multicolumn{1}{|c|}{Humidifier} & \multicolumn{2}{c|}{$\mathtt{N/A}$}  & \multicolumn{2}{c|}{$\mathtt{N/A}$} & \multicolumn{2}{c|}{$\mathtt{N/A}$} & $\mathtt{0.6 \pm 1.4}$ & $\mathtt{1.6 \pm 0.9}$ &\multicolumn{2}{c|}{$\mathtt{N/A}$} & \multicolumn{2}{c|}{$\mathtt{N/A}$}\\ \hline          
          
          \multicolumn{1}{|c|}{Garbage Disposal} & \multicolumn{2}{c|}{$\mathtt{N/A}$} & \multicolumn{2}{c|}{$\mathtt{N/A}$} & $\mathtt{0}$ & $\mathtt{0.2 \pm 0.4}$ & \multicolumn{2}{c|}{$\mathtt{N/A}$} & \multicolumn{2}{c|}{$\mathtt{N/A}$} & \multicolumn{2}{c|}{$\mathtt{N/A}$} \\ \hline
          
          \multicolumn{1}{|c|}{TV} & \multicolumn{2}{c|}{$\mathtt{N/A}$} & $\mathtt{0}$ & $\mathtt{18.7 \pm 37}$ & $\mathtt{0}$ & $\mathtt{0}$ & \multicolumn{2}{c|}{$\mathtt{N/A}$} & \multicolumn{2}{c|}{$\mathtt{N/A}$} & \multicolumn{2}{c|}{$\mathtt{N/A}$} \\ \hline
                  
          \multicolumn{1}{|c|}{Vacuum Robot} & \multicolumn{2}{c|}{$\mathtt{N/A}$} & \multicolumn{2}{c|}{$\mathtt{N/A}$} & \multicolumn{2}{c|}{$\mathtt{N/A}$} & \multicolumn{2}{c|}{$\mathtt{N/A}$} & $\mathtt{0.26 \pm 0.07}$  & $\mathtt{0}$ & \multicolumn{2}{c|}{$\mathtt{N/A}$} \\ \hline          
          
          \multicolumn{1}{|c|}{Light Bulb} & \multicolumn{2}{c|}{$\mathtt{N/A}$}  & $\mathtt{0}$ & $\mathtt{20.2 \pm 4.3}$ & \multicolumn{2}{c|}{$\mathtt{N/A}$} & \multicolumn{2}{c|}{$\mathtt{N/A}$} &\multicolumn{2}{c|}{$\mathtt{N/A}$} & \multicolumn{2}{c|}{$\mathtt{N/A}$}\\ \hline
          
          \multicolumn{1}{|c|}{AC}& $\mathtt{9.6 \pm 3.4}$ & $\mathtt{0}$ &  \multicolumn{2}{c|}{$\mathtt{N/A}$} & $\mathtt{0}$ & $\mathtt{0.4 \pm 0.5}$ & $\mathtt{3.2 \pm 2}$ & $\mathtt{1.2 \pm 1.1}$ & \multicolumn{2}{c|}{$\mathtt{N/A}$} & \multicolumn{2}{c|}{$\mathtt{N/A}$} \\ \hline
          
          \multicolumn{1}{|c|}{Door Lock} & \multicolumn{2}{c|}{$\mathtt{N/A}$} & \multicolumn{2}{c|}{$\mathtt{N/A}$} & $\mathtt{0}$ & $\mathtt{0}$ & \multicolumn{2}{c|}{$\mathtt{N/A}$} & \multicolumn{2}{c|}{$\mathtt{N/A}$} & \multicolumn{2}{c|}{$\mathtt{N/A}$} \\ \hline

        \end{tabular}
        {\large{
        \textbf{Measurement units:} $\mathtt{\tau}$: minutes; $\mathtt{\epsilon}$: depends on the physical quantity; Motion, smoke, sound: binary; Illuminance: luminous flux; Temperature: $\mathtt{\degree F}$; Humidity: $\mathtt{RH\%}$.}}
    \end{threeparttable}}
\end{table}

%% file: 0-CCS2022-FinalVersion/tables/apps_experiment.tex
\begin{table}[t!]
\caption{Apps in our experiments and their descriptions.}
\label{tab:apps}
\resizebox{\columnwidth}{!}{
\begin{threeparttable}
\begin{tabular}{|l|l|}
\hline
\textbf{ID}       & \textbf{Description}                                                                   \\ \hline \hline
$\mathtt{App_{1}}$     & Activates the heater at a user-defined time                                        \\ \hline
$\mathtt{App_{2}}$     & Activates the oven at a user-defined time                                                   \\ \hline
$\mathtt{App_{3}}$     & Activates the pressure cooker at a user-defined time                                        \\ \hline
$\mathtt{App_{4}}$     & Activates the coffee maker at a user-defined time                                         \\ \hline
$\mathtt{App_{5}}$     & Activates the dehumidifier at a user-defined time                                          \\ \hline
$\mathtt{App_{6}}$     & Activates the clothes washer at a user-defined time                                         \\ \hline
$\mathtt{App_{7}}$     & Activates the dryer at a user-defined time                                               \\ \hline
$\mathtt{App_{8}}$     & Activates the humidifier at a user-defined time                                           \\ \hline
$\mathtt{App_{9}}$    & Activates the garbage disposal at a user-defined time                                     \\ \hline
$\mathtt{App_{10}}$    & Activates the TV at a user-defined time                                                     \\ \hline
$\mathtt{App_{11}}$    & Activates the robot vacuum at a user-defined time                                            \\ \hline
$\mathtt{App_{12}}$     & Activates the light bulb at a user-defined time                                           \\ \hline
$\mathtt{App_{13}}$     & Activates the AC at a user-defined time                                              \\ \hline
$\mathtt{App_{14}}$     & Locks the door at a user-defined time                                              \\ \hline
$\mathtt{App_{15}}$    & Turns on ``away mode''  at a user-defined time                                        \\ \hline
$\mathtt{App_{16}}$    & Turns on ``sleep mode''  at a user-defined time                                        \\ \hline

$\mathtt{App_{17}}$    & Activates robot vacuum when the home mode is set to ``away'' or ``vacation''                     \\ \hline
$\mathtt{App_{18}}$    & Turns on the light when sound is detected                                            \\ \hline
$\mathtt{App_{19}}$    & Calls a user's phone when Nest detects sound and the home mode is away                                    \\ \hline
$\mathtt{App_{20}}$    & Turns on a Hubitat device (TV) when motion or sound is detected                \\ \hline
\multirow{2}{*}{$\mathtt{App_{21}}$}   & Activates the heater when the sensor reports too cold; \\ 
&Turns off the heater when temperature is normal                         \\ \hline
$\mathtt{App_{22}}$    & Turns on Kasa plug (humidifier) when temperature is above a threshold          \\ \hline
$\mathtt{App_{23}}$    & Turns on the light when temperature is above or below a threshold          \\ \hline
$\mathtt{App_{24}}$    & Calls the user when motion is detected and home mode is away                                     \\ \hline
$\mathtt{App_{25}}$    & Turns on the light when motion is detected                                        \\ \hline
$\mathtt{App_{26}}$    & Turns on the heater when motion is detected                                        \\ \hline

$\mathtt{App_{27}}$    & Unlocks the door when motion is detected \\ \hline
$\mathtt{App_{28}}$    & Calls the user when smoke is detected                                          \\ \hline
$\mathtt{App_{29}}$    & Turns on all lights when smoke is detected                                  \\ \hline
$\mathtt{App_{30}}$    & Turns on AC if humidity rises above a threshold                                \\ \hline
\multirow{1}{*}{$\mathtt{App_{31}}$}    & Activates the humidifier when the air is dry \\ \hline                        
$\mathtt{App_{32}}$    & Switches off a SmartLife plug (AC) when humidity level is high  \\ \hline
$\mathtt{App_{33}}$    & Flashes a light if humidity level is high                                     \\ \hline
$\mathtt{App_{34}}$    & Turns off the light when the room gets bright                                  \\ \hline
$\mathtt{App_{35}}$    & Turns off humidifier when humidity has risen back to normal level                         \\ \hline
$\mathtt{App_{36}}$    & Turns off dehumidifier when humidity level is low                          \\ \hline

$\mathtt{App_{37}}$    & Turns on/off the AC based on the temperature of the room           \\ \hline
$\mathtt{App_{38}}$    & Turns on the light when the room gets dark                                     \\ \hline
$\mathtt{App_{39}}$    & Turns on the AC when the home mode is set to sleep                         \\ \hline

\end{tabular}
\end{threeparttable}
}
\end{table}

%% file: 0-CCS2022-FinalVersion/tables/actuator_details.tex
\begin{table}[t!]
    \centering
    \caption{Details of the actuators in the house.} 
    \label{tab:actuator_detail}
    \setlength{\tabcolsep}{1.1em}
    \resizebox{\columnwidth}{!}{
    \begin{threeparttable}
        \begin{tabular}{|c|c|c|c|} \hline
        \multirow{2}{*}{\textbf{Device (ID)}} & \textbf{Operating}  & \multirow{2}{*}{\textbf{Distance (m)$^\dagger$}} & \multirow{2}{*}{\textbf{Details$^\ddagger$}} \\
        & \textbf{Time (mins)} &  & \\ \hline
        \multirow{2}{*}{\circled{1}} & \multirow{2}{*}{$\mathtt{set(temp)}$} & \texttt{temp} - $2.8$ & $\mathtt{set(70-82 \degree F)}$ \\ 
        & & \texttt{hum} - $2.8$ & $\mathtt{Dep}$ \\ \hline
        
        \multirow{2}{*}{\circled{2}} & \multirow{2}{*}{$10$} & \texttt{temp} - $1.5$  & $\mathtt{surfTemp = 104 \degree F}$\\ 
        & & \texttt{hum} - $1.5$ & $\mathtt{Dep}$ \\ \hline
        
        \multirow{2}{*}{\circled{3}} & \multirow{2}{*}{$15$} & \texttt{temp} - $0.9$ & $\mathtt{surfTemp = 104 \degree F}$ \\ 
        & & \texttt{hum} - $0.9$ & $\mathtt{Dep}$ \\ \hline 
        
        \multirow{2}{*}{\circled{4}} & \multirow{2}{*}{$3$} & \texttt{temp} - $0.5$ & $\mathtt{surfTemp = 150 \degree F}$ \\ 
        & & \texttt{hum} - $0.5$ & $\mathtt{Dep}$  \\ \hline
        
        \circled{5} & $20$ & \texttt{hum} - $1.5$ & $\mathtt{vaporRem = 0.5 g/min}$ \\ \hline
        
        \multirow{2}{*}{\circled{6}} & \multirow{2}{*}{$25$} & \texttt{hum} - $1.7$& $\mathtt{vaporGen = 0.1g/min}$ \\ 
        & & \texttt{sound} - $1.8$ & $55$ dB at $1$ m \\ \hline
        
        \multirow{3}{*}{\circled{7}} & \multirow{3}{*}{$30$} & \texttt{temp} - $1.8$ & $\mathtt{surfTemp = 96 \degree F}$ \\ 
        & & \texttt{hum} - $1.8$ & $\mathtt{Dep}$ \\ 
        & & \texttt{sound} - $1.8$ & $58$ dB at $1$ m \\ \hline

        \circled{8} & $20$ & \texttt{hum} - $1.8$ & $\mathtt{vaporGen = 0.8g/min}$ \\ \hline
        
        \circled{9} & $0.2$ & \texttt{sound} - $0.8$ & $58$ dB at $1$ m\\ \hline
        
        \multirow{2}{*}{\circled{\scriptsize{10}}} & \multirow{2}{*}{$25$} & \texttt{illum} - $1.2$ & $400$ lumens\\
        & & \texttt{sound} - $2$ & $62$ dB at $1$ m \\ \hline
        
        \circled{\scriptsize{11}} & $20$ & \texttt{motion}  - variable$^*$ & Move towards sensor \\ \hline
        
        \circled{\scriptsize{12}} & $15$ & \texttt{illum}  - $1.4$ & $815$ lumens \\ \hline
        
        \multirow{3}{*}{\circled{\scriptsize{13}}} & \multirow{3}{*}{$\mathtt{set(temp)}$} & \texttt{temp} - $2$ & $\mathtt{set(70-82 \degree F)}$ \\ 
        & & \texttt{hum} - $2$ & $\mathtt{vaporRem = 10g/min}$ \\ 
        & & \texttt{sound} - $3.5$ & $62$ dB at $1$ m \\ \hline
        
        \circled{\scriptsize{14}} & $0.1$ & \texttt{sound} - $3.7$ & $50$ dB at $1$ m\\ \hline
        
        \end{tabular}
        $\dagger$ \texttt{temp}: Temperature Sensor, \texttt{hum}: Humidity Sensor, \texttt{illum}: Illuminance Sensor, \texttt{motion}: Motion Sensor, \texttt{sound}: Sound Sensor\\
        $\ddagger$ $\mathtt{surfTemp}$: Surface Temperature, $\mathtt{vaporGen}$: Water Vapor Generation Rate, $\mathtt{vaporRem}$: Water Vapor Removal Rate, $\mathtt{Dep}$: The actuator impacts the sensor due to dependency \\
        $*$ Robot vacuum moves towards the motion sensor. \\
    \end{threeparttable}}
\end{table}